\documentstyle[12pt]{article}

\newcommand{\om}{\omega}
\newcommand{\al}{\alpha}
\newcommand{\ep}{\epsilon}

\newcommand{\la}{\lambda}

\newcommand{\lb}{\lbrack}
\newcommand{\rb}{\rbrack}

\newcommand{\msc}[1]{\mbox{\scriptsize #1}}
\newcommand{\dsp}{\displaystyle}

\newcommand{\bc}{\mbox{{\bf C}}}
\newcommand{\br}{\mbox{{\bf R}}}
\newcommand{\bz}{\mbox{{\bf Z}}}

\newcommand{\bone}{\mbox{{\bf 1}}}

\newcommand{\bsz}{\msc{{\bf Z}}}

\newcommand{\da}{\dot{a}}
\newcommand{\dal}{\dot{\alpha}}
\newcommand{\db}{\dot{b}}

\newcommand{\cA}{{\cal A}}
\newcommand{\cB}{{\cal B}}

\newcommand{\cK}{{\cal K}}

\newcommand{\cJ}{{\cal J}}

\newcommand{\cO}{{\cal O}}

\newcommand{\cN}{{\cal N}}

\newcommand{\cP}{{\cal P}}
\newcommand{\cF}{{\cal F}}
\newcommand{\cQ}{{\cal Q}}
\newcommand{\cH}{{\cal H}}

\newcommand{\ket}[1]{{|#1\rangle}}

\newcommand{\lsim}{\stackrel{<}{\sim}}
\newcommand{\gsim}{\stackrel{>}{\sim}}

\newcommand{\nn}{\nonumber\\}

\newcommand {\eqn}[1]{(\ref{#1})}

\makeatletter
\@addtoreset{equation}{section}
\def\theequation{\thesection.\arabic{equation}}
\makeatother

\begin{document}
\vskip 7mm

\begin{titlepage}
 \
 \renewcommand{\thefootnote}{\fnsymbol{footnote}}
 \font\csc=cmcsc10 scaled\magstep1
 {\baselineskip=14pt
 \rightline{
 \vbox{\hbox{hep-th/0205200}
       \hbox{UT-02-31}
       }}}

 \vfill
 \baselineskip=20pt
 \begin{center}
 \centerline{\Huge  Superstrings on PP-Wave Backgrounds} 
 \vskip 5mm 
 \centerline{\Huge   and Symmetric Orbifolds}

 \vskip 2.0 truecm
\noindent{\it \large Yasuaki Hikida and Yuji Sugawara} \\
{\sf hikida@hep-th.phys.s.u-tokyo.ac.jp~,~
sugawara@hep-th.phys.s.u-tokyo.ac.jp}
\bigskip

 \vskip .6 truecm
 {\baselineskip=15pt
 {\it Department of Physics,  Faculty of Science, \\
  University of Tokyo \\
  Hongo 7-3-1, Bunkyo-ku, Tokyo 113-0033, Japan}
 }
 \vskip .4 truecm

 \end{center}

 \vfill
 \vskip 0.5 truecm

\begin{abstract}
\baselineskip 6.7mm

We study the superstring theory on pp-wave background with NSNS-flux
that is realized as the Penrose limit of $AdS_3\times S^3 \times M^4$,
where $M^4$ is $T^4$ or $T^4/\bz_2(\cong K3)$. 
Quantizing this system in the covariant gauge, we explicitly construct
the space-time supersymmetry algebra and the complete set of DDF
operators. 
We analyse the spectrum of physical states 
by using the spectrally flowed representations of current algebra.
This spectrum is  classified by the ``short string sectors'' 
and the ``long string sectors'' as in $AdS_3$ string theory. 
The states of the latter propagate freely along the transverse plane of
pp-wave background, but the states of the former do not. 
We compare the short string spectrum with the BPS and almost BPS states 
which have large R-charges in the symmetric orbifold conformal theory, 
which is known as the candidate of dual theory of superstrings on 
$AdS_3\times S^3 \times M^4$.  We show that every short string states 
can be embedded successfully in the single particle Hilbert space 
of symmetric orbifold conformal theory. 

\end{abstract}

\vfill

\setcounter{footnote}{0}
\renewcommand{\thefootnote}{\arabic{footnote}}
\end{titlepage}
\baselineskip 18pt


\newpage

\section{Introduction}
\indent

Recently the string theories on pp-wave backgrounds have been studied
intensively. They are realized by the Penrose limits of 
the string theories on $AdS$ spaces \cite{BFHP} and 
it has just discovered that the string theory on pp-wave background with
RR-flux can be quantized in the light-cone gauge \cite{Metsaev}.
The authors of \cite{BMN} applied these facts to the $AdS/CFT$
correspondence \cite{AdSCFT} and they have made a remarkable progress beyond
the supergravity approximation. 
More precisely, they considered  the ``almost BPS
states'' with large R-charges, which slightly break supersymmetry, 
and succeeded in constructing the single trace operators describing the
string excitations in the large $N$ supersymmetric Yang-Mills theory. 
Many developments have been presented since then.
The orbifoldizations \cite{orbifold} and the extensions including 
D-branes \cite{open} have been investigated widely 
and the other recent developments are given in 
\cite{RT,Hatsuda,recent,Berkovits,KP,interact,PS}.

The cases of the Penrose limits of $AdS_3\times S^3$ are somewhat special
because we can also consider the NSNS-flux.
Such backgrounds are exactly soluble as noncompact WZW models 
associated with the 6-dimensional Heisenberg group ($H_6$),
which are simple generalizations of the Nappi-Witten models \cite{NW}.
Contrary to the cases with RR-flux\footnote
  {The quantization of the case with RR-flux in the covariant gauge 
  is discussed recently in \cite{Berkovits}.},
we can quantize these systems covariantly by using the standard 
NSR formalism  of superstring theory.

The superstring theory on $AdS_3\times S^3 \times M^4$ 
($M^4 =T^4$ or $K3$) with NSNS-flux is realized as the near horizon 
limit of NS1-NS5 system \cite{GKS}. 
This is a system of great importance because
it gives a possibility to make analysis beyond 
the supergravity approximation. 
The famous candidate of the dual theory 
is the 2-dimensional $\cN=(4,4)$ non-linear $\sigma$-model 
on the symmetric orbifold $Sym^{Q_1Q_5}(M^4) 
\equiv (M^4)^{Q_1Q_5}/S_{Q_1Q_5}$ \cite{Maldacena,MS}. 
($Q_1$ and $Q_5$ are the NS1 and NS5 charges, respectively.) 
There are many attempts to understand this duality from the world-sheet
picture of string theory (e.g., \cite{GKS,DORT,KS}). 
One of the important tests of this duality is the comparison of
the  spectrum of BPS states. The BPS states with small R-charges 
are explicitly identified  \cite{KLL,HS},
and the higher R-charged BPS states are also 
reproduced successfully in \cite{HHS,AGS}
with the help of the spectral flow symmetry \cite{HHRS,MO}. 
However, the analysis beyond the BPS spectrum
has been difficult and still incomplete.

In order to overcome this difficulty we shall follow the idea of \cite{BMN}.
More precisely, we will concentrate on the states which possess 
very large R-charges and break the BPS condition slightly. 
These states are often called  as ``almost BPS states''.
The sectors of such states are well described by taking the Penrose limit. 
Under this limit, the string theory on $AdS_3\times S^3 \times M^4$
reduces to the  WZW model on  $H_6 \times M^4$, as we mentioned above.
This model is much easier to treat and the string Hilbert space 
is expected to be completely analysed.

In this paper, motivated by these observations,
we study the superstrings on the pp-wave backgrounds with NSNS-flux,
more precisely, $H_6 \times M^4$ spaces ($M^4 = T^4, ~ T^4/\bz_2$) 
in the covariant gauge quantization. We construct 
the space-time supersymmetry algebra by using the physical string vertices 
in the manner similar to \cite{GKS} and present the complete set 
of DDF operators. {}From this result we investigate 
the spectrum of physical states.  In this analysis the spectrally flowed 
representation plays an essential role just as in the case of $AdS_3$ string. 
The spectrum is classified by the ``short string sectors'' 
and the ``long string sectors'' as in $AdS_3$ string theory \cite{MO}. 
The strings of former sectors cannot propagate along the transverse
plane of pp-wave geometry, while the strings of latter sectors freely
propagate and have a continuous spectrum of light-cone energies. 
Finally we compare the short string spectrum with the 
(almost) BPS states in the conformal field theory on symmetric orbifold. 
We find out successfully the natural embedding 
of all the string states  in the single particle Hilbert space
of symmetric orbifold theory.  
We also comment on the existence of many missing states, which should be
identified with the non-perturbative excitations in string theory side.

~

This paper is organized as follows.
In section 2, which is a preliminary section, 
the space-time supersymmetry (``super pp-wave algebra'') in the $H_6
\times T^4$ background is examined by contracting the supersymmetry
algebra of superstring theory on $AdS_3\times S^3 \times T^4 $. 
In section 3, we quantize the superstring theory on $H_6 \times T^4$
in the covariant gauge and construct the physical vertices 
generating the super pp-wave algebra and the complete set of DDF operators.
{}From this result we analyse the spectrum of physical states explicitly. 
In section 4, starting with a short review of the symmetric orbifold, 
we compare the (almost) BPS spectrum with the string spectrum on the pp-wave
background. We also discuss the extension to the case of 
$H_6 \times T^4/\bz_2 (\cong H_6\times K3)$ in section 5, and 
the section 6 is devoted to conclusion and discussion.

~

\section{Space-time Superalgebra of PP-Wave Background}
\indent

In this preliminary section
we examine the supersymmetric algebra of pp-wave
background by contracting that of  $AdS_3 \times S^3$.
It is well-known that the supersymmetry in the background $AdS_3
\times S^3$ is represented by the super Lie group $PSU(1,1|2) \times
PSU(1,1|2)$. The even part corresponds to the isometry of this background.
The isometry of $AdS_3$ space is identified as $SU(1,1) \times SU(1,1)
(\cong SL(2;\br)\times SL(2;\br))
\sim SO(2,2)$ and the isometry of $S^3$ is identified as $SU(2) \times
SU(2) \sim SO(4)$. 
We denote the generators of $SU(1,1)$ Lie algebra as
$m^{\alpha}_{~\beta}$ $(\alpha , \beta = 1 ,2 )$,
and the generators of
$SU(2)$ Lie algebra as $m^i_{~j}$ $(i,j = \hat{1} ,\hat{2} )$
by following the notation of \cite{MT}. 
(Here, we concentrate on the holomorphic sector and
the anti-holomorphic sector can be defined in the similar way.)
The commutation relations are given by 
\begin{equation}
 {[} m^{\alpha}_{~\beta} , m^{\gamma}_{~\delta} {]}= 
 \delta^{\gamma}_{~\beta} m^{\alpha}_{~\delta} - 
 \delta^{\alpha}_{~\delta} m^{\gamma}_{~\beta} ~,~
 {[} m^i_{~j} , m^k_{~n} {]}= 
 \delta^k_{~j} m^i_{~n} - \delta^i_{~n} m^{k}_{~j} ~,
\end{equation}
and the Hermitian conjugations are defined as
\begin{eqnarray}
&& (m^1_{~1})^{\dagger} = m^1_{~1} ~,~
 (m^2_{~1})^{\dagger} = m^1_{~2} ~,~
 (m^1_{~2})^{\dagger} = m^2_{~1} ~, \nn
&& (m^{\hat{1}}_{~\hat{1}})^{\dagger} = m^{\hat{1}}_{~\hat{1}} ~,~
 (m^{\hat{2}}_{~\hat{1}})^{\dagger} = - m^{\hat{1}}_{~\hat{2}} ~,~
 (m^{\hat{1}}_{~\hat{2}})^{\dagger} = - m^{\hat{2}}_{~\hat{1}} ~.
\end{eqnarray} 
The generators of odd sector $q^{\alpha}_{~i}$ and $q_{~\alpha}^{i}$ 
correspond to $8 (+ 8)$ supercharges and the commutation relations are
\begin{eqnarray}
&& {[} m^{\alpha}_{~\beta} , q^k_{~\gamma}  {]}= 
 - \delta^{\alpha}_{~\gamma} q^k_{~\beta} 
 + \frac{1}{2}  \delta^{\alpha}_{~\beta} q^k_{~\gamma} ~,~~
 {[} m^i_{~j} , q^k_{~\alpha} {]} =
  \delta^k_{~j} q^i_{~\alpha} 
 - \frac{1}{2}  \delta^i_{~j} q^k_{~\alpha} ~, \nn
&& {[} m^i_{~j} , q^{\alpha}_{~k} {]} = 
 - \delta^i_{~k} q^{\alpha}_{~j} 
 + \frac{1}{2}  \delta^i_{~j} q^{\alpha}_{~k} ~,~~
 {[} m^{\alpha}_{~\beta} , q^{\gamma}_{~k} {]} = 
  \delta^{\gamma}_{~\beta} q^{\alpha}_{~k} 
 - \frac{1}{2}  \delta^{\alpha}_{~\beta} q_{~k}^{\gamma} ~, \nn
&& \{ q^i_{~\alpha} , q^{\beta}_{~j}\} =
 i (\delta^i_{~j} m^{\beta}_{~\alpha} + \delta^{\beta}_{~\alpha} m^i_{~j})~.
\end{eqnarray}
The Hermitian conjugations are defined as 
\begin{eqnarray}
&& (q^{1}_{~\hat{1}})^{\dagger} = - i q^{\hat{1}}_{~1} ~,~
 (q^{2}_{~\hat{2}})^{\dagger} = - i q^{\hat{2}}_{~2} ~,~
 (q^{2}_{~\hat{1}})^{\dagger} = i q^{\hat{1}}_{~2} ~,~
 (q^{1}_{~\hat{2}})^{\dagger} = i q^{\hat{2}}_{~1} ~, \nn
&& (q_{~1}^{\hat{1}})^{\dagger} = i q_{~\hat{1}}^{1} ~,~
 (q_{~2}^{\hat{2}})^{\dagger} = i q_{~\hat{2}}^{2} ~,~
 (q_{~1}^{\hat{2}})^{\dagger} = - i q_{~\hat{2}}^{1} ~,~
 (q_{~2}^{\hat{1}})^{\dagger} = - i q_{~\hat{1}}^{2} ~.
\end{eqnarray}

Now, we take a light-cone basis and a Penrose limit. 
First, we consider the even sector. 
We redefine the generators as 
\begin{eqnarray}
 J &=& - m^1_{~1} + m^{\hat{1}}_{~\hat{1}} ~,~~
 F ~=~ - \frac{1}{R} \left( m^1_{~1} + m^{\hat{1}}_{~\hat{1}} \right) ~, \nn
 P_1 &=&  \frac{i}{\sqrt{R}} m^2_{~1} ~,~~ 
 P_1^*  ~=~ - \frac{i}{\sqrt{R}} m^1_{~2} ~, \nn
 P_2 &=& -  \frac{i}{\sqrt{R}} m^{\hat{1}}_{~\hat{2}} ~,~~ 
 P_2^*  ~=~ - \frac{i}{\sqrt{R}} m^{\hat{2}}_{~\hat{1}} ~,
\end{eqnarray}
and take the limit of $R \to \infty$. 
Then we obtain the commutation relations as
\begin{equation}
 {[} J , P_i {]} = P_i ~,~~ 
 {[} J , P_i^* {]} = - P_i^* ~,~~ 
 {[} P_i , P_j^* {]} = \delta_{ij} F ~,
\label{CR1}
\end{equation}
which is the same as the ones of $H_6$ Lie algebra. 
The Hermitian conjugations are given by
\begin{eqnarray}
 (J)^{\dagger} = J ~,~
 (F)^{\dagger} = F ~,~
 (P_i)^{\dagger} = P_i^* ~,~
 (P_i^*)^{\dagger} = P_i ~.
\end{eqnarray} 

The analysis of odd part can be done just like the even part.
We redefine the generates of odd sector as 
\begin{eqnarray}
&& Q^{--+} = - \frac{i}{\sqrt{R}} q^1_{~\hat{1}} ~,~~ 
   Q^{+++} = -  \frac{i}{\sqrt{R}} q^2_{~\hat{2}} ~, \nn
&& Q^{++-} = - \frac{1}{\sqrt{R}} q_{~1}^{\hat{1}} ~,~~ 
   Q^{---} =  - \frac{1}{\sqrt{R}} q_{~2}^{\hat{2}} ~, \nn
&& Q^{-++} = q^1_{~\hat{2}} ~,~~ 
   Q^{+-+} =  q^2_{~\hat{1}} ~, \nn
&& Q^{-+-} = i q_{~2}^{\hat{1}} ~,~~ 
   Q^{+--} = i q_{~1}^{\hat{2}} ~,
\end{eqnarray}
and take the limit of $R \to \infty$.
The commutation relations with the generators of even sector becomes 
\begin{eqnarray}
&& {[} J , Q^{+ + a} {]} =  Q^{+ + a}  ~,~~
 {[} J , Q^{- - a} {]} = -  Q^{- - a}  ~, \nn
&& {[} P_1 , Q^{- + a} {]} = - Q^{+ + a}  ~,~~
 {[} P_1^* , Q^{+ - a} {]} =  Q^{- - a}  ~, \nn
&& {[} P_2 , Q^{+ - a} {]} = - Q^{+ + a}  ~,~~
 {[} P_2^* , Q^{- + a} {]} = - Q^{- - a}  ~, 
\label{CR2}
\end{eqnarray}
and the other commutation relations vanish.
The anti-commutation relations among odd sector are obtained as
\begin{eqnarray}
&& \{ Q^{--a} ,Q^{++b} \} = \epsilon^{ab} F ~,~~
   \{ Q^{-+a} ,Q^{+-b} \} = \epsilon^{ab}  J ~, \nn
&& \{ Q^{++a} ,Q^{+-b} \} =  \epsilon^{ab}  P_1 ~,~~
   \{ Q^{-+a} ,Q^{--b} \} =   \epsilon^{ab} P_1^* ~,\nn
&& \{ Q^{-+a} ,Q^{++b}\} = \epsilon^{ab} P_2 ~,~~
   \{ Q^{+-a} ,Q^{--b} \} = \epsilon^{ab} P_2^* ~.
\label{CR3}
\end{eqnarray}
The Hermitian conjugations are given by
$ (Q^{\epsilon_1, \epsilon_2, \epsilon_3})^{\dagger} = 
 Q^{-\epsilon_1, -\epsilon_2, -\epsilon_3}$.
The authors of \cite{Hatsuda} obtained  the superalgebras of the
pp-wave backgrounds by contracting the
superalgebras of $AdS_5 \times S^5$,  $AdS_4 \times S^7$ and $AdS_7
\times S^4$. 
Our result is the counterpart of the case of $AdS_3 \times S^3$ and
we obtain the supersymmetric extension of $H_6$ Lie algebra.

~

\section{Superstring Theory on PP-Wave Background with NSNS-Flux}
\indent

The superstring theory on the $AdS_3 \times S^3 (\times M^4) $ ($M^4 = T^4$
or $K3$) with NSNS-flux can be described by the $SL(2;\br) \times SU(2)$ 
super WZW model.  The Penrose limit is realized by contracting 
these currents \cite{Sfetsos} and it becomes the WZW model 
whose target space is 6 dimensional Heisenberg group $H_6$. 
This kind of models were investigated in \cite{KK,NW2,RT2} and 
studied recently with newer motivations \cite{RT,KP,PS}. 
The sigma model approach of quantization was taken in \cite{RT2,RT}
and the current algebra method was developed  
in \cite{KK,KP} by using free field realization. 
Although these two approaches should be equivalent, 
we shall use the latter method, because the similarity with
the analysis  developed in \cite{GKS}
for the case of $AdS_3 \times S^3$ becomes more transparent.

\subsection{$H_6$ Super WZW Model as a Penrose Limit}
\indent

The superstring theory on the $AdS_3 \times S^3$ can be described by the
$SL(2;\br) \times SU(2)$ super WZW model. We set the level of 
each current algebra to an equal value  $k \in \bz_{>0}$.  
(The bosonic parts of current algebras  $SL(2;\br)$ and $SU(2)$ 
have the levels $k+2$ and $k-2$, respectively).  
This system is described by the super current algebras 
\begin{eqnarray}
&&\cJ^A(z,\theta) = \sqrt{\frac{k}{2}}\psi^A(z)
+\theta J^A(z)~ (\mbox{for $SL(2;\br)$})~,\nn
&&\cK^a(z,\theta) = \sqrt{\frac{k}{2}}\chi^a(z)
+\theta K^a(z)~ (\mbox{for $SU(2)$})~.
\label{super SL(2) SU(2)}
\end{eqnarray}
The ``total currents''
$J^A(z)$ and $K^a(z)$ $(A,a = \pm , 3)$
satisfy the following operator product expansions (OPEs)
\begin{eqnarray}
&& J^3 (z) J^3 (w) \sim - \frac{k}{2(z-w)^2} ~,~~
 J^3 (z) J^{\pm} (w) \sim \frac{\pm J ^{\pm}(w)}{z-w} ~, \nn
&& J^+ (z) J^- (w) \sim \frac{k}{(z-w)^2} - \frac{2 J^3 (w)}{z-w} ~, \\
&& K^3 (z) K^3 (w) \sim \frac{k}{2(z-w)^2} ~,~~
 K^3 (z) K^{\pm} (w) \sim \frac{\pm K ^{\pm}(w)}{z-w} ~, \nn
&& K^+ (z) K^- (w) \sim \frac{k}{(z-w)^2} + \frac{2 K^3 (w)}{z-w} ~,
\end{eqnarray}
and the free fermions $\psi^A$ and $\chi^a$ are defined by the OPEs
\begin{eqnarray}
&&\psi^3(z)\psi^3(w) \sim -\frac{1}{z}~,~~~ 
\psi^+(z)\psi^-(w) \sim \frac{2}{z-w}~,\nn
&&\chi^3(z)\chi^3(w) \sim \frac{1}{z}~,~~~ 
\chi^+(z)\chi^-(w) \sim \frac{2}{z-w}~,
\end{eqnarray}
and transformed by the action
of the total currents as follows
\begin{eqnarray}
&& J^3(z)\psi^{\pm}(w) \sim \psi^3(z)J^{\pm}(w) \sim 
\frac{\pm \psi^{\pm}(w)}{z-w}~, \nn
&& J^{\pm}(z)\psi^{\mp}(w) \sim \mp \frac{2\psi^3(w)}{z-w}~, \\
&& K^3(z)\chi^{\pm}(w) \sim \chi^3(z)K^{\pm}(w) \sim 
\frac{\pm \chi^{\pm}(w)}{z-w}~, \nn
&& K^{\pm}(z)\chi^{\mp}(w) \sim \pm \frac{2\chi^3(w)}{z-w}~,
\end{eqnarray}

The Penrose limit of this model is given by 
a noncompact super WZW model associated with
the 6 dimensional Heisenberg group $H_6$, which is a natural
generalization of the  Nappi-Witten model \cite{KK,NW2,RT2}.
According to \cite{Sfetsos}, 
we can obtain the supercurrents of this model by contracting
those of the $SL(2;\br)\times SU(2)$ model.
We redefine the  supercurrents \eqn{super SL(2) SU(2)} as 
\begin{eqnarray}
 \cJ(z,\theta) &=& \cK^3(z,\theta) + \cJ^3(z,\theta) ~,~~ 
 \cF(z,\theta) ~=~ \frac{1}{k} (\cK^3(z,\theta) - \cJ^3(z,\theta)) ~, \nn
 \cP_1(z,\theta)  &=& \frac{1}{\sqrt{k}} \cJ^+(z,\theta) ~,~~
 \cP_1^*(z,\theta)  ~=~ \frac{1}{\sqrt{k}} \cJ^-(z,\theta) ~, \nn
 \cP_2(z,\theta)  &=& \frac{1}{\sqrt{k}} \cK^+(z,\theta) ~,~~
 \cP_2^*(z,\theta)  ~=~ \frac{1}{\sqrt{k}} \cK^-(z,\theta) ~,
\label{contraction}
\end{eqnarray}
and take the limit $k \to \infty$ with keeping the eigenvalues of
$K^3_0-J^3_0$ order $\cO(k)$ but the eigenvalues of 
$K^3_0+J^3_0$ much smaller than $k$.
Then we obtain the supercurrent algebra of the  $H_6$ super WZW model as
\begin{eqnarray}
&& {\cal J} (\theta , z) = \psi_J (z) + \theta J (z) ~,~~
 {\cal F} (\theta , z) = \psi_F (z) + \theta F (z) ~, \nn
&& {\cal P}_i (\theta , z) = \psi_{P_i} (z) + \theta P_i (z) ~,~~
 {\cal P}_i^* (\theta , z) = \psi_{P_i^*} (z) + \theta P_i^* (z) ~,
\end{eqnarray}
where $i = 1,2$.
The total currents $J(z)$, $F(z)$, $P_i(z)$ and $P^*_i(z)$ 
satisfy the OPEs
\begin{eqnarray}
 J(z) P_i (w) &\sim& \frac{P_i (w)}{z-w} ~,~~
 J(z) P^*_i (w) ~\sim~ - \frac{P^*_i (w)}{z-w} ~, \nn
 P_i (z) P_j^* (w) &\sim& \delta_{ij} 
 \left( \frac{1}{(z-w)^2} + \frac{F(w)}{z-w}\right) ~,\nn
 J (z) F (w) &\sim& \frac{1}{(z-w)^2}~.
\label{H6OPE}
\end{eqnarray}
Other OPEs have no singular terms. 
Their superpartners $\psi_J$, $\psi_F$, $\psi_{P_i}$ and $\psi_{P_i^*}$ 
are free fermions defined by
\begin{equation}
 \psi_{P_i} (z) \psi_{P_j^*} (w) \sim \frac{\delta_{ij}}{z-w} ~,~~
 \psi_J (z) \psi_F (w) \sim \frac{1}{z-w} ~.
\end{equation}
The total currents $J$, $F$, $ P_i$ and $P_i^*$ 
non-trivially act on these free fermions as
\begin{eqnarray}
&& J (z) \psi_{P_i} (w) \sim \psi_J(z)P_i(w) \sim
\frac{\psi_{P_i}}{z-w} ~,\nn
&& J (z) \psi_{P^*_i} (w) \sim \psi_J(z)P^*_i(w) \sim
- \frac{\psi_{P^*_i}}{z-w} ~,\nn
&& P_i (z) \psi_{P_j^*} (w) \sim \psi_{P_i}(z)P_j^*(w) \sim
\delta_{ij} \frac{\psi_{F}}{z-w} ~.
\end{eqnarray}

The physical meaning of this contraction is clear. 
In the context of $AdS_3/CFT_2$ correspondence, the eigenvalue of $-J^3_0$ 
corresponds to the conformal weight $\Delta$ 
(and also the space-time energy)
in the boundary conformal theory and the eigenvalue of $K^3_0$ 
is interpreted as the space-time R-charge $Q$.
Therefore we can expect that the above contraction 
amounts to focusing on the states of string theory on $AdS_3 \times S^3$  
(or the boundary conformal theory) that are very close to the BPS bound
and have large R-charges.
We call them as ``almost BPS states'' according to \cite{BMN} and 
they are characterized more precisely as
\begin{equation}
  \Delta+Q \sim k\gg 1~, ~~~ \Delta-Q \ll k ~.
\end{equation}


The $\cN=1$ superconformal symmetry is realized as follows.
Because the total currents have non-trivial OPEs with fermions, it is
convenient to introduce the ``bosonic currents'' which can be treated
independently of fermions. They are defined as
\begin{eqnarray}
&& \hat{J}  = J - \psi_{P_1} \psi_{P_1^*} - \psi_{P_2} \psi_{P_2^*} ~,~~
   \hat{F} =  F ~, \nn
&& \hat{P}_i = P_i - \psi_F \psi_{P_i} ~,~~
   \hat{P}_i^* = P_i^* + \psi_F \psi_{P_i^*} ~,
\end{eqnarray} 
which again satisfy the same OPE \eqn{H6OPE} but have no singular OPEs
with the free fermions.
The $\cN=1$ supercurrent is now defined in the standard fashion
\begin{eqnarray}
 G &=& \hat{J} \psi_F + \hat{F} \psi_J 
 + \sum_{i=1}^2 (\hat{P}_i \psi_{P_i^*} + \hat{P}_i^* \psi_i 
    + \psi_F \psi_{P_i} \psi_{P_i^*} )  \nonumber \\
  &\equiv&   J \psi_F + F \psi_J 
 + \sum_{i=1}^2 (\hat{P}_i \psi_{P_i^*} + \hat{P}_i^* \psi_i) ~,
\end{eqnarray}
and the total currents can be given by acting $G$ on the fermions
$\psi_J$, $\psi_F$, $\psi_{P_i}$ and $\psi_{P_i^*}$ as it should be.

The analysis can be performed  by using the abstract current algebra
techniques in principle.   However, it is easier to make use of the 
following free field realizations as given in  \cite{KK}.
We introduce the free bosons $X^+$, $X^-$, $Z_i$ and $Z_i^*$ $(i=1,2)$
defined by the OPEs
\begin{equation}
 X^+ (z) X^- (w) \sim  - \ln (z-w) ~,~~ 
 Z_i (z) Z_j^* (w) \sim - \delta_{ij} \ln (z-w) ~,
\end{equation}
and  rewrite the fermions $\psi_J$, $\psi_F$, $\psi_{P_i}$ and
$\psi_{P_i^*}$ as 
\begin{equation}
 \psi_F = \psi^+ ~,~~\psi_J = \psi^- ~,~~
 \psi_{P_i} = \psi_i e^{i X^+} ~,~~ \psi_{P_i^*} = \psi_i^* e^{-iX^+} ~,
\end{equation}
where the new fermions are defined by
\begin{equation}
 \psi^+ (z) \psi^- (w) \sim \frac{1}{z-w} ~,~~
 \psi_i (z) \psi_j^* (w) \sim  \frac{\delta_{ij}}{z-w} ~.
\end{equation}
The total currents can be now expressed as 
\begin{eqnarray}
&& F = i \partial X^+ ~,~~ J = i \partial X^- ~, \nn
&& P_i = e^{i X^+} ( i \partial Z_i + \psi^+ \psi_i) ~,~~
  P_i^* = e^{- i X^+} ( i \partial Z_i^* - \psi^+ \psi_i^*) ~,
\end{eqnarray}
and the bosonic currents are written as
\begin{eqnarray}
&& \hat{F} = i \partial X^+ ~,~~ 
  \hat{J} = i \partial X^- - \psi_1 \psi_1^* - \psi_2 \psi_2^* 
   - 2 i \partial X^+~, \nn
&& \hat{P}_i = e^{i X^+}  i \partial Z_i  ~,~~
 \hat{P}_i^* = e^{- i X^+} i \partial Z_i^*  ~.
\end{eqnarray}
In terms of these free fields
the superconformal current is rewritten as the standard form
of flat background
\begin{equation}
 G =  \psi^{+} i \partial X^- +  \psi^{-} i \partial X^+ 
     + \psi_i^* i \partial Z_i + \psi_i i \partial Z^*_i  ~.
\end{equation} 
We also introduce the free fields  $Y^i$ and $\la^i$ $(i=1,2,3,4)$
to describe the remaining  $T^4$ sector as 
\begin{equation}
 \lambda^i (z) \lambda^j (w) \sim \frac{\delta^{ij}}{z-w} ~,~~  
 Y^i(z)Y^j(w) \sim -\delta^{ij}\ln (z-w)~.
\end{equation}

In order to analyze the space-time fermions, we have to introduce the spin
fields, which are defined by using the bosonized fermions.
We bosonize the fermions as
\begin{eqnarray}
 \psi^{\pm 0} &=& \psi^{\pm} ~=~  e^{\pm i H_0}~, \nn
 \psi^{+ j} &=& \psi_j ~=~ e^{+ i H_j} ~, \nn
 \psi^{- j} &=&  \psi_j^* ~=~ e^{- i H_j} ~, \nn
 \psi^{\pm 3} &=&
 \frac{1}{\sqrt{2}}(\lambda^1 \pm i \lambda^2 ) = e^{\pm i H_3}~,\nn 
 \psi^{\pm 4} &=&
 \frac{1}{\sqrt{2}}(\lambda^3 \pm i \lambda^4 ) = e^{\pm i H_4}~,
\end{eqnarray}
and define the spin fields as
\begin{equation}
 S^{\epsilon_0 \epsilon_1 \epsilon_2 \epsilon_3 \epsilon_4} 
 = \exp \left(  \frac{i}{2} \sum_{j=0}^{4} \epsilon_j H_j \right) ~.
\label{spin field}
\end{equation}
The GSO condition imposes $ \prod_{j=0}^4\ep_j= + 1$ in the
convention of this paper.
Precisely speaking, 
the OPEs including the spin fields are affected by the cocycle factors
and they depend on the notation of gamma matrices. 
We summarize our convention of gamma matrices in appendix A.


\subsection{Hilbert Space of $H_6$ Super WZW Model}
\indent

The irreducible representations of the current algebra
of Nappi-Witten model (that is the $H_4$ WZW model)
are classified in \cite{KK}. It is easy to extend to the case 
of $H_6$ super WZW model. We shall here focus on 
the ``type II'' representation (corresponding to the highest weight 
representation of the zero-mode subalgebra) in the terminology of
\cite{KK}, and later we discuss the other types of representations. 
The vacuum state (in the NS sector) is characterized by
\begin{eqnarray}
&& J_0 \ket{j,\eta} = j \ket{j,\eta} ~,~~
 F_0 \ket{j,\eta} =  \eta  \ket{j,\eta} ~,\nn
&& P_{i,n} \ket{j,\eta} = 0 ~,~({}^\forall n \geq 0) ~,~~
 P_{i,n}^* \ket{j,\eta} = 0 ~,~({}^\forall n > 0) ~, \nn
&& \Psi_r \ket{j,\eta} = 0 ~,~({}^\forall r  > 0~, ~ r\in \frac{1}{2}+\bz)~ ,
\label{type II}
\end{eqnarray}
where $\Psi$ represents all the fermionic fields $\psi_J$, $\psi_F$,  
$\psi_{P_i}$ and $\psi_{P_i^*}$. We assume that $j\in \br$ and $0<\eta <1$.

It is useful to rewrite this representation \eqn{type II} 
in terms of the free fields $X^{\pm}$, $Z_i$, $Z_i^*$, $\psi^{\pm}$,
$\psi_i$ and $\psi_i^*$. 
This is nothing but a Fock representation 
with the Fock vacuum defined by the vertex operator
\begin{equation}
 V = \exp \left( i j X^+ + i\eta X^-\right) \sigma_{\eta} ~,  
\end{equation}
where $\sigma_{\eta}$ is the (chiral) twist field. 
This field imposes the boundary conditions
\begin{eqnarray}
&& i\partial Z_i(e^{2\pi i}z) = e^{-2\pi i \eta} i\partial Z_i(z)~,~~
\psi_i(e^{2\pi i}z) = e^{-2\pi i \eta} \psi_i(z)~, \nn
&& i\partial Z^*_i(e^{2\pi i}z) = e^{2\pi i \eta} i\partial Z^*_i(z)~,~~
\psi^*_i(e^{2\pi i}z) = e^{2\pi i \eta} \psi^*_i(z)~,
\end{eqnarray}
which ensure the locality of $H_6$ supercurrents.
More precisely, $\sigma_{\eta}$ should have the following OPEs
\begin{eqnarray}
&& i \partial Z_i (z) \sigma_\eta (w) \sim 
 (z-w)^{- \eta} {\tau^i}_\eta (w) ~,~~
 i \partial Z_i^* (z) \sigma_\eta (w) \sim 
 (z-w)^{\eta - 1} {{\tau '}^{i}}_\eta (w) ~, \nn
&& \psi_i (z) \sigma_\eta (w) \sim 
 (z-w)^{- \eta} {t^i}_\eta (w) ~,~~
 \psi^*_i (z) \sigma_\eta (w) \sim 
 (z-w)^{\eta} {{t'}^{i}}_\eta (w) ~, 
\end{eqnarray}
where ${\tau^i}_\eta$, ${{\tau '}^{i}}_\eta$, ${t^i}_\eta$ and
${{t '}^{i}}_\eta$ are the descendant twist fields.  
This twist operator $\sigma_{\eta}$ has the conformal weight
\begin{equation}
h(\sigma_{\eta})= 2 \times \frac{1}{2} \eta (1-\eta) 
 + 2 \times \frac{1}{2} \eta^2 = \eta ~.
\end{equation}

As already discussed in \cite{KK,KP}, there is the spectral flow symmetry 
\begin{eqnarray}
 J_n \to J_n ~,~~ F_n \to F_n + p \delta_{n,0} ~,~~ 
 P_{i,n} \to P_{i,n + p} ~,~~P_{i,n}^* \to P_{i,n-p}^* ~, \\
 \psi_{J,\,r} \to \psi_{J, \, r} ~,~~ \psi_{F,\,r} \to \psi_{F,\,r} ~,~~ 
 \psi_{P_i,\,r} \to \psi_{P_i,\, r + p} ~,~~
 \psi_{P^*_i,\,r} \to \psi_{P^*_i,\,r-p}~,
\label{spectral flow}
\end{eqnarray}
and hence we also consider 
the ``flowed representation'' as the natural extensions
of \eqn{type II}. The vacuum states are given by (where we use $p\in \bz$ as
the spectral flow number)  
\begin{eqnarray}
&& J_0 \ket{j,\eta, p} = j \ket{j,\eta, p} ~,~~
 F_0 \ket{j,\eta, p} =  (\eta+p)  \ket{j,\eta, p} ~,\nn
&& P_{i,n} \ket{j,\eta,p} = 0 ~,~({}^\forall n \geq -p) ~,~~
 P_{i,n}^* \ket{j,\eta,p} = 0 ~,~({}^\forall n > p) ~, \nn
&& \psi_{J,\,r} \ket{j,\eta,p} =
  0 ~,~({}^\forall r  > 0)~ ,~~
 \psi_{F,\,r} \ket{j,\eta,p} = 
  0 ~,~({}^\forall r  > 0)~ , \nn
&& \psi_{P_i,\,r} \ket{j,\eta,p} = 
  0 ~,~({}^\forall r  > -p)~ ,~~
  \psi_{P^*_i,\,r} \ket{j,\eta,p} = 
  0 ~,~({}^\forall r  > p)~ ,
\label{flowed type II}
\end{eqnarray}
where $n \in \bz$ and $r \in 1/2 + \bz$.
The reason why we should introduce the flowed representations  is 
clear. This spectral flow symmetry \eqn{spectral flow} 
is actually the counterpart of that on $SL(2;\br)\times SU(2)$ WZW
model and it is not difficult to confirm it directly by taking pp-wave
limit \eqn{contraction}.  
In the case of the $AdS_3$ strings, the necessity of flowed
representations is well-known \cite{MO}. 
We will later focus on the sectors with non-zero spectral flow number $p$
to realize the (almost) BPS states and discuss
the correspondence with the symmetric orbifold theory.
We can again realize the flowed representation \eqn{flowed type
II} by means of the Fock representation, in which the Fock 
vacuum corresponds to the vertex operator
\begin{equation}
 V = \exp \left( i j X^+ + i(\eta+p) X^-\right) \sigma_{\eta} ~.  
\end{equation}

~

A few comments are in order:

~

\noindent{ \bf 1.}
We also have other types of irreducible representations
of $H_6$ current algebra.
The ``Type III'' representations have the lowest weights
of zero-mode subalgebra and the eigenvalues of $F$ take $-1 < \eta < 0$. 
The ``Type I'' representations have neither highest
weights nor lowest weights and the eigenvalues of $F$ are $\eta=0$.
There are also their spectrally flowed representations. 
As in the case of $SL(2;\br)$ WZW model, (see, for example, \cite{MO}) 
the spectral flow symmetry interchanges the type III representation with the
type II representation.   
In fact, we can easily see that 
\begin{equation}
\cH^{(\msc{II})}_{j,\eta,p} \cong \cH^{(\msc{III})}_{j,\eta-1,p+1}~,
\label{II III}
\end{equation}
where $\cH^{\msc{(II)}}_{j,\eta,p}$ is the flowed type II
representation defined by using the vacuum \eqn{flowed type II} 
and $\cH^{\msc{(III)}}_{j,\eta,p}$ is the flowed type III representation
defined similarly.  
(We must also make the redefinitions of 
fermionic oscillators as $\psi'_{P_i,\,r}:= \psi_{P_i,\,r+1}$ and 
$\psi'_{P^*_i,\,r}:= \psi_{P^*_i,\,r-1}$ to equate the both sides of 
\eqn{II III}.) Therefore, we only have to consider the type II representation 
if assuming the spectral flow symmetry.  We also note that 
the $p\geq 0$ representations correspond to the positive energy states and 
the $p<0$ representations to the negative energy ones in the original 
$AdS_3$ string theory.

On the other hand, the type I representations seem to be
the counterparts of the principal continuous series in the $AdS_3$
string theory.  
Because there are no twisted fields in this sector, 
the vacua of type I representations simply correspond to the next vertex
operators 
\begin{equation}
 V = \exp \left( ipX^- + 
i ( j + n ) X^+ + i p_i^* Z_i + i p_i Z_i^*\right) ~, ~~ 
 n \in \bz ~.
\end{equation}
If we do not consider the spectral flow $(p=0)$, there are only  
trivial massless states with zero energy,
that do not propagate along the transverse
plane ($Z_i$, $Z_i^*$ directions)\footnote
     {There is no such massless states 
     in the case of principal continuous series in $AdS_3$ strings.
     This fact is due to the existence of the background charge term 
     $Q \sim  \sqrt{\frac{2}{k}}$. However, this term can be 
     neglected under the large $k$ limit. 
     This is the reason why there are such extra massless states.}.
For the cases of flowed type I representations ($p\neq 0$),  many
physical states are possible. 
The spectra of light-cone energies in these sectors are continuous
and the strings freely propagate along the transverse plane. 
It is quite natural to identify these sectors with the ``long string
sectors'' in the $AdS_3$ string \cite{MO}.
On the other hand, the type II (and type III) 
representation should correspond to the ``short string sectors'', 
because they are the counterparts of the discrete series 
in the $AdS_3$ string.   
The strings in these sectors cannot freely propagate 
along the transverse plane because the string coordinates $Z_i$, $Z_i^*$
are twisted.

~

\noindent{ \bf 2.}  We here remark the equivalence
\begin{equation}
\bigoplus_{n\in \bsz}\, \cH^{(\msc{II})}_{j+\frac{n\eta}{p+\eta}, \eta,p}
\cong 
\bigoplus_{n\in \bsz}\, \cF_{j+\frac{n\eta}{p+\eta}, \eta,p}~,
\end{equation}
where we denote the right hand side as the Fock representation of 
free fields $X^{\pm}$, $Z_i$, $Z_i^*$, $\psi^{\pm}$, $\psi_i$ and $\psi_i^*$.
In this sense, we can reproduce the exact physical Hilbert space 
by using the free field representation with no subtlety,
(for example, the treatment of screening charges in the case of $AdS_3$
strings)  
and this fact is a great advantage of taking the Penrose limit.

\subsection{Physical Vertices of Superstring  on PP-Wave Background}
\indent 

Now, we analyse the physical vertex operators. 
We shall concentrate on the short string sectors (type
II representation) with positive energies ($p\geq 0$) for the time
being, and we will discuss later the long string sectors. 
In order to construct the physical vertices, 
it is convenient to make use of 
the free field representation previously discussed.
We fix the Fock vacuum as 
\begin{equation}
\ket{j,\eta,p} = \sigma_{\eta} e^{ijX^+ + i(\eta+p)X^-}
\ket{0}~,~~ 0<\eta<1,~~~ p\in \bz_{\geq 0}~.
\end{equation}
We construct the physical vertex operators explicitly by the so-called
DDF operators in the covariant gauge.
Here we introduce the superghosts $(\gamma, \beta)$ or the
bosonized ones $(\phi, \xi, \eta)$.
The BRST charge has the standard form of the free superstring theory as
\begin{equation}
Q_{\msc{BRST}} = \oint \left\lb c\left(T-\frac{1}{2}(\partial \phi)^2
  -\partial^2\phi -\eta\partial\xi + \partial c b\right)
+\eta e^{\phi}G-b\eta\partial\eta e^{2\phi}\right\rb  ,  \\
\end{equation}
where $T$ and $G$ are the total stress tensor and the superconformal 
current constructed from the free fields $X^{\pm}$, $Z_i$, $Z_i^*$, $Y^i$,
$\psi^{\pm}$, $\psi_i$, $ \psi^*_i$ and $ \la^i$.

The most important vertex operators are the generators
of space-time supersymmetry algebra  \eqn{CR1}, \eqn{CR2} and \eqn{CR3} 
considered above.
The generators of bosonic part are quite easy.
They are nothing but the zero-modes of world-sheet $H_6$ (total) currents;
\begin{eqnarray}
&& \cJ = \oint \psi^- e^{-\phi} 
    =  \oint i \partial X^- =J_0  ~, \nn
&& \cF = \oint \psi^+ e^{-\phi}   
     =   \oint i \partial X^+ =F_0 ~,\nn
&& \cP_i = \oint \psi_i e^{i  X^+} e^{-\phi}  
= \oint \left( i \partial Z_i + \psi^+\psi_i \right)e^{i X^+}= P_{i,\,0}~,\nn
&&  \cP_i^* ~=~ \oint \psi^*_i e^{- i  X^+} e^{-\phi} 
= \oint\left( i \partial Z^*_i 
- \psi^+\psi_i^* \right) e^{- i  X^+} = P_{i,\,0}^*  ~,
\label{PVF}
\end{eqnarray}
where we implicitly identify the operators by using the picture changing
operator. 
The generators of fermionic part are obtained by using the 
spin fields (\ref{spin field}) as (in the $(-1/2)$ picture)
\begin{eqnarray}
&& \cQ^{++a} = \oint S^{+++aa} e^{i X^+} e^{- \frac{\phi}{2}} ~,~~ 
\cQ^{--a} =  \oint S^{+--aa} e^{-i X^+} e^{- \frac{\phi}{2}} ~,\nn 
&& \cQ^{+-a} = \oint S^{-+-aa}  e^{- \frac{\phi}{2}} ~,~~
\cQ^{-+a} =  \oint S^{--+aa} e^{- \frac{\phi}{2}} ~.
\label{PVS}
\end{eqnarray}
These operators manifestly BRST invariant and 
they can locally act on the Fock space associated with
$\ket{j,\eta,p}$ (irrespective of the values of $j$, $\eta$ and $p$).
We can directly check that they generate the super pp-wave 
algebra (\ref{CR1}), (\ref{CR2}) and (\ref{CR3}). 
In particular, we note that
\begin{eqnarray}
&& \{\cQ^{-+a}, \cQ^{+-b} \} = \ep^{ab}\cJ~, ~~~ 
\lb \cJ, \cQ^{\pm\mp a}\rb  =0~, 
\end{eqnarray}
which indicates that $\cQ^{-+a}$ and $\cQ^{+-a}$ play the role of 
supercharges with the ``Hamiltonian'' $\cJ$.

In order to analyse the spectrum of physical states
we further need to introduce the DDF operators.
Recalling the analysis in $AdS_3$ string theory \cite{GKS},
it is quite natural to consider the ``affine extension'' 
of \eqn{PVF} and \eqn{PVS} as
\begin{eqnarray}
\cP_{i,\,n} &=& \frac{1}{\sqrt{p+\eta}}
\oint \psi_i e^{i\frac{n+\eta}{p+\eta}X^+}e^{-\phi}~, \nn
\cP^*_{i,\,n} &=&  \frac{1}{\sqrt{p+\eta}}
\oint \psi_i e^{i\frac{n-\eta}{p+\eta}X^+}e^{-\phi}~,\nn
\cQ^{++a}_n &=&  \frac{1}{\sqrt{p+\eta}}
\oint S^{+++aa} e^{i\frac{n+\eta}{p+\eta}X^+}
 e^{- \frac{\phi}{2}} ~,\nn 
\cQ^{--a}_n &=&   \frac{1}{\sqrt{p+\eta}}
\oint S^{+--aa}  e^{i\frac{n-\eta}{p+\eta}X^+}
 e^{- \frac{\phi}{2}}~.
\label{DDF1}
\end{eqnarray}
These operators are BRST invariant and locally act on the Fock space 
associated with $\ket{j,\eta,p}$ (of the fixed $\eta$ and $p$). 
It is obvious that
\begin{eqnarray}
&&\sqrt{p+\eta}\cP_{i,\,p}= \cP_i ~,~~ 
\sqrt{p+\eta}\cP^*_{i,\,-p}= \cP^*_i ~, \nn
&&\sqrt{p+\eta}\cQ^{++a}_p = \cQ^{++a} ~,~~ 
\sqrt{p+\eta}\cQ^{--a}_{-p}=\cQ^{--a} ~.
\end{eqnarray}
Note that the supercharges $\cQ^{\pm \mp a}$
do not have such affine extensions because the BRST invariance cannot 
be preserved. The DDF operators \eqn{DDF1} satisfy  
the following (anti-)commutation relations 
(up to the picture changing and BRST exact terms)
\begin{eqnarray}
&&\lb \cP_{i,\,m}, \cP^*_{j,\,n}\rb = \frac{m+\eta}{p+\eta}
\delta_{ij}\delta_{m+n,0}~,~~~
\{ \cQ_m^{--a}, \cQ_n^{++b} \} =  \ep^{ab}\delta_{m+n,0}~, \nn
&&\lb \cJ , \cP_{i,\,n} \rb =  \frac{n+\eta}{p+\eta}\cP_{i,\,n}~, ~~~
\lb \cJ , \cP^*_{i,\,n} \rb =  \frac{n-\eta}{p+\eta}\cP^*_{i,\,n}~, \nn
&&\lb \cJ , \cQ^{++a}_{n} \rb =  \frac{n+\eta}{p+\eta}\cQ^{++a}_{n}~, ~~~
\lb \cJ , \cQ^{--a}_{n} \rb =  \frac{n-\eta}{p+\eta}\cQ^{--a}_{n}~, 
\label{DDF comm 1}
\end{eqnarray}
It is also useful to remark that $(\cP_{i,\,n}, \cQ^{++a}_n)$ and
$(\cP^*_{i,\,n}, \cQ^{--a}_n)$ are the supermultiplets 
with respect to the supercharges $\cQ^{+-a}$ and $\cQ^{-+a}$.
More precisely, we find the relations
\begin{eqnarray}
  \cP_{1,\,n}  &\stackrel{\cQ^{-+a}}{\longrightarrow}&
  \cQ_n^{++a}  ~\stackrel{\cQ^{-+(-a)}}{\longrightarrow}~
  \cP_{2,\,n} ~\stackrel{\cQ^{-+*}}{\longrightarrow}~  0 ~ \nn
  0~ \stackrel{\cQ^{+-*}}{\longleftarrow}~ 
  \cP_{1,\,n}  &\stackrel{\cQ^{+-(-a)}}{\longleftarrow}&
  \cQ_n^{++a} ~\stackrel{\cQ^{+-a}}{\longleftarrow}~
  \cP_{2,\,n} ~,
\label{CRPQ}
\end{eqnarray}
\begin{eqnarray}
  \cP^*_{1,\,n}  &\stackrel{\cQ^{+-a}}{\longrightarrow}&
  \cQ_n^{--a}  ~\stackrel{\cQ^{+-(-a)}}{\longrightarrow}~
  \cP^*_{2,\,n} ~\stackrel{\cQ^{+-*}}{\longrightarrow}~  0 ~ \nn
  0~ \stackrel{\cQ^{-+*}}{\longleftarrow}~ 
  \cP^*_{1,\,n}  &\stackrel{\cQ^{-+(-a)}}{\longleftarrow}&
   \cQ_n^{--a} ~\stackrel{\cQ^{-+a}}{\longleftarrow}~
  \cP^*_{2,\,n} ~.
\label{CRP*Q}
\end{eqnarray}
The explicit forms of the (anti-)commutation relations are summarized in
appendix B.

In order to construct the remaining DDF operators for the $T^4$ 
directions, it is convenient to relabel the  fermions $\la^i$ as 
\begin{eqnarray}
 \lambda^{+-} &=& \frac{1}{\sqrt{2}} (\lambda^1 + i \lambda^2 ) ~,~~ 
 \lambda^{-+} ~=~ \frac{1}{\sqrt{2}} ( - \lambda^1 + i \lambda^2 ) ~,\nn
 \lambda^{++} &=& \frac{1}{\sqrt{2}} ( - \lambda^3 - i \lambda^4 ) ~,~~ 
 \lambda^{--} ~=~ \frac{1}{\sqrt{2}} ( - \lambda^3 + i \lambda^4 ) ~,
\end{eqnarray}
and the free bosons $Y^{a\da}$ are defined similarly.
These have the following OPEs 
\begin{equation}
 \lambda^{a \dot{a}} (z) \lambda^{b \dot{b}} (w) \sim 
  \frac{ \epsilon^{ab} \epsilon^{\dot{a} \dot{b}} }{z-w} ~, ~~~
 Y^{a\da}(z)Y^{b\db}(w) \sim - \ep^{ab}\ep^{\da\db} \ln (z-w)~. 
\end{equation}
The DDF operators can be then given by
\begin{eqnarray}
 \cA^{a \dot{a}}_n &=& \frac{1}{\sqrt{p+\eta}}
 \oint \lambda^{a \dot{a}} e^{i \frac{n}{p + \eta}  X^+} e^{-\phi} 
=  \frac{1}{\sqrt{p+\eta}}
 \oint i\partial Y^{a \dot{a}} e^{i \frac{n}{p + \eta}  X^+}  
~, \nn
 \cB^{+ \pm}_n &=& - \frac{1}{\sqrt{p+\eta}} \oint S^{+ - + \mp \pm}  
 e^{i \frac{n}{p+\eta}  X^+} e^{-\frac{\phi}{2}} ~, \nn
 \cB^{- \pm}_n &=&  \frac{1}{\sqrt{p+\eta}} \oint S^{+ + - \mp \pm}  
 e^{i \frac{n}{p + \eta}  X^+} e^{-\frac{\phi}{2}} ~,
\label{DDF2}
\end{eqnarray}
and they satisfy  
\begin{eqnarray}
&& {[}  \cA^{a \dot{a}}_m , \cA^{b \dot{b}}_n {]}
  = \frac{m}{p+\eta} \delta_{m+n} \epsilon^{ab} \epsilon^{\dot{a} \dot{b}} ~,~~
 {[}  \cB^{\al \dot{a}}_m , \cB^{\beta \dot{b}}_n {]}
  =  \delta_{m+n} \epsilon^{ab} \epsilon^{\dot{a} \dot{b}} ~,\nn
&&\lb \cJ,  \cA^{a \dot{a}}_n \rb = \frac{n}{p+\eta}\cA^{a \dot{a}}_n~,~~~
\lb \cJ,  \cB^{\al \dot{a}}_n \rb = \frac{n}{p+\eta}\cB^{\al \dot{a}}_n~.
\label{DDF comm 2}
\end{eqnarray}
The pair of $(\cA^{a\da}_n, \cB^{\al\da}_n)$ again become supermultiplets 
with respect to $\cQ^{\pm \mp a}$ (see appendix B for the more precise
relations)
\begin{eqnarray}
  \cA^{a\da}_n  &\stackrel{\cQ^{\mp \pm (-a)}}{\longrightarrow}&
  \cB^{\pm \da}_n  ~\stackrel{\cQ^{\pm \mp(-a)}}{\longrightarrow}~
  \cA^{(-a)\da}_n ~\stackrel{\cQ^{**(-a)}}{\longrightarrow}~  0 ~ \nn
  0~ \stackrel{\cQ^{**a}}{\longleftarrow}~ 
  \cA^{a\da}_n   &\stackrel{\cQ^{\pm\mp a}}{\longleftarrow}&
  \cB^{\pm \da}_n  ~\stackrel{\cQ^{\mp \pm a}}{\longleftarrow}~
  \cA^{(-a)\da}_n ~.
\label{CRAQ}
\end{eqnarray}

Finally we discuss whether or not
the other vertex operators can  be constructed
from the remaining spin fields $S^{-\pm\pm\pm\mp}$ and
$S^{-\pm\pm\mp\pm}$, which are allowed by  the GSO condition.
The locality condition requires the ``dressing'' 
of the factors like $e^{\pm i\frac{n+\eta}{p+\eta}X^+}$, 
however these factors are not compatible with the BRST invariance. 
Therefore, we conclude that 
$(\cP_{i,\,n}, \cQ^{++a}_n)$, $(\cP^*_{i,\,n}, \cQ^{--a}_n)$ and 
$(\cA^{a\da}_n, \cB^{\al\da}_n)$ are the complete set of DDF
operators.

\subsection{Spectrum of Physical States}
\indent

Because we have presented the complete list of DDF operators, it 
is now easy to construct general physical states. 
More precisely speaking, we are interested in  the physical states 
corresponding to the almost BPS states in the original 
$AdS_3\times S^3$ superstring theory characterized by 
\begin{equation}
     \Delta + Q \sim k \gg 1~, ~~~ \Delta-Q \ll k ~,
\label{BPS condition}
\end{equation}
where $\Delta$ represents the space-time energy, (measured by $-J^3_0$ 
in the original $AdS_3$ string theory) that is identified as the 
conformal weight of the boundary conformal field theory, 
and $Q$ represents the space-time R-charge (measured by $K^3_0$).  
This condition is equivalent to\footnote
   {In our convention, the BPS inequality is equivalent to $\cJ \leq 0$.}   
\begin{equation}
\cF \gsim 1~, ~~~|\cJ| \ll k ~,
\end{equation}
and all string excitations satisfy this condition  
in the pp-wave limit $k \to + \infty $. 
The BPS states correspond to the cases of $\cJ=0$ and belong to  
the short multiplets of the superalgebra \eqn{PVF} and \eqn{PVS}.
The condition $\cF \gsim 1$ only leads to the restriction $p\geq 1$
with respect to the spectral flow number $p$.
(we only consider the positive energy states.)

We concentrate on  the sectors with no momenta along $T^4$ direction
for the time being. 
It is not difficult to write down the complete list of BPS states for
each of the fixed $p$ and $\eta$. 
In the NS sector, we obtain (where we focus on the left-mover only)
\begin{eqnarray}
&& \ket{\omega^0 ; \eta, p} 
= \psi_{1,-\frac{1}{2} + \eta} \ket{0,\eta,p} 
    \otimes c e^{-\phi} \ket{0}_{\msc{gh}} ~, \nn
&& \ket{\omega^2 ; \eta, p}  = \psi_{2,-\frac{1}{2} + \eta} \ket{0,\eta,p} 
    \otimes c e^{-\phi} \ket{0}_{\msc{gh}} ~, 
\label{BPS NS}
\end{eqnarray}
and we have two more BPS states in the R-sector as
\begin{eqnarray}
&& \ket{\omega^{1 \pm} ; \eta, p} =
  (S^{-++\mp \pm})_{-\frac{5}{8} + \eta } \ket{0,\eta,p} 
 \otimes c e^{-\frac{\phi}{2}} \ket{0}_{\msc{gh}} ~.
\label{BPS R}
\end{eqnarray}
Here, we point out the next relation, which is useful for our discussion
\begin{eqnarray}
  \ket{\omega^0;\eta,p}  &\stackrel{\cB_0^{+\dot{a}}}{\longrightarrow}&
  \ket{\omega^{1\dot{a}};\eta,p}  
 ~\stackrel{\cB_0^{+(-\dot{a})}}{\longrightarrow}~
  \ket{\omega^2;\eta,p} ~\stackrel{\cB_0^{+ *}}{\longrightarrow}~  0 ~ \nn
  0~ \stackrel{\cB_0^{- *}}{\longleftarrow}~ 
  \ket{\omega^0;\eta,p}  &\stackrel{\cB_0^{-(-\dot{a})}}{\longleftarrow}&
  \ket{\omega^{1\dot{a}};\eta,p}~\stackrel{\cB_0^{- \dot{a}}}{\longleftarrow}~
  \ket{\omega^2;\eta,p} ~.
\label{BPS relation}
\end{eqnarray}
Now, there is an obvious correspondence  with the ``chiral part'' of 
the cohomology ring of $T^4$ by identifying $\cB^{+\da}_0$ with 
the holomorphic one-form $dZ^{\da}$ on $T^4$. 
Therefore, emphasizing the correspondence to $H^*(T^4)$,
the non-chiral BPS states can be explicitly written as
\begin{equation}
\ket{\om^{(q,\bar{q})}; \eta, p} = \ket{\om^q;\eta,p}\otimes 
  \overline{\ket{\om^{\bar{q}};\eta,p}} ~,~~~
({}^{\forall} \om^{(q,\bar{q})} \in H^{q,\bar{q}}(T^4)) ~. 
\end{equation}
These states have the degenerate charge $\cF = p+\eta$ 
for each sector of $p$ and $\eta$.

We can construct the other types of physical states by making  
the DDF operators \eqn{DDF1} and \eqn{DDF2} act on these BPS states.
We first note that the BPS states are actually the Fock vacua 
with respect to \eqn{DDF1} and \eqn{DDF2}; that is
\begin{eqnarray}
&& \cP_{i,\,n}\ket{\om;\eta,p}=0~,~({}^{\forall}n\geq 0)~,~~~
\cP^*_{i,\,n}\ket{\om;\eta,p}=0~,~({}^{\forall}n>0)~,  \nn
&& \cQ_n^{++a}\ket{\om;\eta,p}=0~,~({}^{\forall}n\geq 0)~, ~~~
\cQ_n^{--a}\ket{\om;\eta,p}=0~,~({}^{\forall}n> 0)~, \nn
&& \cA^{a\da}_n\ket{\om;\eta,p}=0~,~({}^{\forall}n\geq 0)~,~~~
\cB^{\al\da}_n\ket{\om;\eta,p}=0~,~({}^{\forall}n>0)~.
\end{eqnarray}
(For $\cB^{\al\da}_0$, see \eqn{BPS relation}.)
We  hence  obtain  
\begin{eqnarray}
&&  \cB_{-n_1}^{\alpha_1 \dot{a}_1} \cdots 
  \cQ_{-m_1}^{++b_1} \cdots 
  \cQ_{-k_1}^{--c_1} \cdots
  \otimes
  \bar{\cB}_{-\bar{n}_1}^{\bar{\al}_1 \bar{\dot{a}}_1} \cdots 
  \bar{\cQ}_{-\bar{m}_1}^{++\bar{b}_1} \cdots 
  \bar{\cQ}_{-\bar{k}_1}^{--\bar{c}_1} \cdots
\ket{\om; \eta, p} ~,\nn
&& \hspace{1.7cm} n_i, \bar{n}_i, m_i, \bar{m}_i >0~,~ k_i, \bar{k}_i \geq 0~,~
 {}^{\forall}\om \in H^*(T^4)~,
\label{almost BPS1}
\end{eqnarray} 
as typical physical states.
These states become almost BPS under the condition
\begin{eqnarray}
&& |\cJ+\bar{\cJ}| \equiv  
  \frac{1}{p+\eta} \left\lb 
\left(\sum_i n_i + \sum_i (m_i-\eta) + \sum_i (k_i+\eta) \right) 
 \right.\nn
&& \hspace{2.5cm} +  \left. \left(\sum_i \bar{n}_i + \sum_i (\bar{m}_i-\eta) 
+ \sum_i (\bar{k}_i+\eta)\right) \right\rb  \ll  k  ~,
\end{eqnarray}
which is always satisfied for sufficiently large $k$.
Other states can be obtained by multiplying the supercharges 
$\cQ^{\pm\mp a}$.
In order to consider the level matching condition, 
we must keep it in mind that
our free field representation corresponds to choosing different coordinate
systems for the left and right movers as discussed in \cite{KP}. 
Hence we can include the non-vanishing ``helicity in the transverse plane''
$\cJ-\bar{\cJ} = h\in \bz$, and the level matching condition becomes
\begin{eqnarray}
&&\left(\sum_i n_i + \sum_i (m_i-\eta) + \sum_i (k_i+\eta)\right)
 \nn && \hspace{1.7cm} 
-\left(\sum_i \bar{n}_i + \sum_i (\bar{m}_i-\eta) 
+ \sum_i (\bar{k}_i+\eta) \right)   \in (p+\eta)\bz~.
\label{lm string}
\end{eqnarray}

We should note that under the limiting procedure $k \to \infty$, 
huge number of stringy excitations of the original $AdS_3$
superstring theory are included in our physical Hilbert space of string
theory on pp-wave background.
These states could correspond to very massive 
states, (which could possess very large energies $-J^3_0$) 
although they have the small $\cJ$-charges.  
This fact gives us a theoretical ground for making it possible 
to identify a lot  of stringy excitations with the objects 
in the dual theory as in \cite{BMN}.


In order to complete our discussion
we must also  consider the sectors with non-trivial momenta
along $T^4$. It is easy to show that there are no BPS states in these
sectors. However, it is possible to construct the almost BPS states. 
We only consider a rectangular torus for simplicity and use $R_a$ 
$(a=1,2,3,4)$ as the radii. 
The momenta of $T^4$ sector can be written as 
\begin{equation}
p_a = \frac{n_a}{R_a}+\frac{w^aR_a}{2}~,~~
\bar{p}_a = \frac{n_a}{R_a}-\frac{w^aR_a}{2}~,~~ (n_a, w^a \in \bz)~,
\label{T4 momenta}
\end{equation}
where $n_a$ and $w^a$ are the KK momenta and winding modes,
respectively. Here we use the convention $\al'\equiv l_s^2=2$.
The simplest physical states (in NSNS sector) have the next form
\begin{equation}
\psi_{i,-\frac{1}{2} + \eta}\bar{\psi}_{\bar{i},-\frac{1}{2} + \eta} 
\ket{j,\bar{j},\eta,p;n_a,w^a} 
    \otimes c\bar{c} e^{-\phi-\bar{\phi}} \ket{0}_{\msc{gh}} ~,
\label{phys T4 momenta}
\end{equation}
where $\ket{j,\bar{j},\eta,p;n_a,w^a}$ corresponds to the vertex
operator 
\begin{equation}
e^{i(jX^++(p+\eta)X^-+ p_aY^a)}\otimes 
e^{i(\bar{j}\bar{X}^++(p+\eta)\bar{X}^-+ \bar{p}_a\bar{Y}^a)} ~.
\end{equation}
The on-shell condition leads to 
\begin{equation}
j = -\frac{1}{2(p+\eta)}\sum_a p_a^2~, ~~~
\bar{j} = -\frac{1}{2(p+\eta)}\sum_a \bar{p}_a^2~,
\end{equation} 
and the level matching condition $j-\bar{j} \in \bz $ amounts to 
\begin{equation}
 \sum_a n_aw^a \in (p+\eta)\bz  ~.
\end{equation}
The general physical states are obtained by making the DDF operators
\eqn{DDF1} and \eqn{DDF2} and supercharges $\cQ^{\pm \mp a}$ act 
on the above states \eqn{phys T4 momenta}. 
The condition for the almost BPS states is again given by  
\begin{equation}
|\cJ+\bar{\cJ}| \ll k~,
\end{equation} 
and the level matching condition is 
\begin{equation}
 \cJ -\bar{\cJ} \in \bz~.
\end{equation}
We have again huge number of stringy states
for sufficiently large $k$.


~

Finally we make a few comments:

~

\noindent{\bf 1.}
The spectrum of light-cone energies is given by 
$H_{\msc{l.c.}} = -(\cJ+\bar{\cJ})$.
In particular, 
in the case of $p=0$ (although we are interested in the cases 
of $p\geq 1$), we can obtain 
\begin{equation}
H_{\msc{l.c.}} = -(\cJ+\bar{\cJ}) = \frac{1}{\eta}(\mbox{N}+\bar{\mbox{N}}) 
  + \mbox{J}+\bar{\mbox{J}} 
+ \frac{1}{2\eta}\left(\sum_ap_a^2+\sum_a\bar{p}_a^2\right)  ~,  
\end{equation}
where $\mbox{N}$ and  $\bar{\mbox{N}}$ are the mode counting  operators 
and $\mbox{J}$ and $\bar{\mbox{J}}$ are the ``angular momentum
operators''. They act on the DDF operators as 
\begin{eqnarray}
&& \lb \mbox{N}, \cO_n\rb = -n \cO_n~,~~~
(\cO_n = \cP_{i,\,n}, \cP^*_{i,\,n}, \cQ^{++a}_n, \cQ^{--a}_n, 
\cA^{a\da}_n, \cB^{\al\da}_n) ~,\nn
&& \lb \mbox{J}, \cO_n\rb = - \cO_n~,~~~ 
(\cO_n = \cP_{i,\,n},\cQ^{++a}_n)~,\nn
&&\lb \mbox{J}, \cO_n\rb = \cO_n~,~~~ 
(\cO_n = \cP^*_{i,\,n},\cQ^{--a}_n)~,\nn
&&\lb \mbox{J}, \cO_n\rb = 0~,~~~ 
(\cO_n = \cA^{a\da}_n, \cB^{\al\da}_n)~.
\end{eqnarray}
The level matching condition is expressed as 
$\cJ-\bar{\cJ} \equiv h \in \bz$, which leads 
to the conditions $\mbox{N}-\bar{\mbox{N}}=0$ and
$\mbox{J}-\bar{\mbox{J}}\equiv h \in \bz$ 
for generic value of $\eta$, as discussed in \cite{KP}.
This spectrum is consistent with the result given in \cite{RT}
(and the appendix of \cite{BMN}).

~

\noindent{\bf 2.}
We can also construct many physical states in the sectors 
of spectrally flowed type I representations. 
The analysis is quite easy because it can be described by the usual free
fields without twist operators.
As we have already mentioned, it is plausible to suppose that 
they correspond to the long strings in the $AdS_3$ string theory
\cite{MMS,SW,MO}. 
More precisely, the strings in these sectors possess a continuous 
spectrum of the light-cone energies
and can freely propagate along the transverse plane. 

The corresponding excitations do not seem to exist in the symmetric
orbifold theory, which we will analyse in the next section, because 
it only includes the discrete spectrum. 
The existence of such continuous spectrum in the pp-wave strings 
reflects the non-compactness of the background, 
and is presumably related to the singularity of the type discussed in
\cite{SW}.  On the other hand, the symmetric orbifold theory corresponds to the
``smooth point'' in the moduli space of NS1-NS5 system with the
non-vanishing world-sheet theta angle.  {}From this reason 
the absence of the continuous spectrum does not indicate a
contradiction. The detailed analysis of such singularity and 
the aspects of long string sectors  will be significant subjects
for our future study, 
however we shall not discuss them  any more in this paper.

~


\section{Comparison with  $Sym^{M}(T^4)$ $\sigma$-Model }
\indent

The well-known candidate of the dual theory of the superstrings on the  
$AdS_3 \times S^3 \times T^4$ is the $\cN=(4,4)$ 
non-linear $\sigma$-model on the symmetric orbifold space 
$Sym^M(T^4) \equiv (T^4)^M/S_M$, where $M= Q_1Q_5 $, for sufficiently
large $Q_1$ and $Q_5$ \cite{Maldacena,MS}. 
($Q_1$ and $Q_5$ are the NS1 and NS5 charges, respectively.) 
The charge $Q_5$ can be identified as the level $k$ of $SL(2;\br)$ 
and $SU(2)$ WZW models.
The charge $Q_1$ should be the implicit
upper bound of spectral flow number $p$. 
(See, for example, \cite{KS}. Of course, $p$ should not be 
bounded from the viewpoints of perturbative string theory.
However, we take anyway the large $Q_1$ limit as well as large $Q_5$.)
The main subject of this section is to analyse the spectrum of BPS and
almost BPS states with large R-charges $Q (\gsim Q_5 \gg 1)$, 
and compare the spectrum of short string sectors with positive energies,
which was studied in the previous section. 
To this aim, we start with a short review of $Sym^M(T^4)$  $\sigma$-Model.

\subsection{Short Review of $Sym^{M}(T^4)$ $\sigma$-Model}
\indent

The $\cN=(4,4)$ superconformal field theory defined by the
supersymmetric $\sigma$-model on symmetric orbifold 
$Sym^{M}(T^4)\equiv (T^4)^M/S_M$  is described as follows.  
We use  $4M$  free bosons 
$X^{a\da}_{(A)}$ $(A=0,1,2,\ldots, M-1, a, \da = \pm)$ and free fermions
$\Psi^{\al\da}_{(A)}$ and $\bar{\Psi}^{\dal\da}_{(A)}$ 
 as the fundamental fields.
The superconformal symmetry is realized by the following currents
(where we only write the left-mover)
\begin{eqnarray}
&&  T(z) = -\frac{1}{2}\sum_{A}\, \ep_{ab}\, \ep_{\da\db} \, 
           \partial X_{(A)}^{a\da} \, \partial X_{(A)}^{b\db} 
          -\frac{1}{2}\sum_{A} \,  \ep_{\al\beta}\, \ep_{\da\db}\,
             \Psi_{(A)}^{\al\da}\, \partial \Psi_{(A)}^{\beta\db} ~, \nn
&& G^{\al a}(z) = i\sum_{A}\, \ep_{\da\db}\, 
    \Psi_{(A)}^{\al\da} \, \partial X_{(A)}^{a\db} ~, \nn
&& K^{\al\beta}(z) = - \frac{1}{2}\, \sum_{A} \, \ep_{\da\db} \,
            \Psi_{(A)}^{\al\da} \, \Psi_{(A)}^{\beta\db} ~.
\label{SCA}
\end{eqnarray}
These  generate the $N=4$ (small) superconformal algebra (SCA)
with central charge $c=6M$.
In our convention, we set $\ep^{+-}=\ep_{-+}=1$ and $\ep^{-+}=\ep_{+-}=-1$,
and the OPEs of free fields are written as 
\begin{eqnarray}
X_{(A)}^{a\da}(z)X_{(B)}^{b\db}(0) \sim 
-\delta_{AB}\ep^{ab}\ep^{\da\db}\ln z  ~, ~~
\Psi^{\al\da}(z)\Psi^{\beta\db}(0) \sim 
\delta_{AB}   \frac{ \ep^{\al\beta}\ep^{\da\db} }{z} ~.
\end{eqnarray}
The usual convention of $SU(2)$ current is given by 
$K^3 = K^{+-}= K^{-+}$ and $K^{\pm}=\pm K^{\pm \pm}$.
The subalgebra of ``zero-modes'' $\{ L_{\pm 1},\, L_0, \, 
G_{\pm 1/2}^{\al a}, \, K_0^{\al\beta} \}$ and the counterpart of the 
right mover compose the super Lie algebra
$PSU(1,1|2)_L \times PSU(1,1|2)_R$.
This subalgebra  corresponds to the
supersymmetric algebra on the $AdS_3\times S^3$ geometry,
as we already mentioned.

According to the general approach to the orbifold conformal field theory 
\cite{orbifold CFT}, 
we have various twisted sectors corresponding to the each element $g$ of 
$S_{M}$. The Hilbert space of each twisted sector is defined 
with the following boundary condition $(z\equiv e^{\tau +i\sigma})$ as
\begin{equation}
\Phi_{(A)}(\tau, \sigma +2\pi ) = \Phi_{g(A)}(\tau, \sigma) ~, 
\label{twisted bc}
\end{equation}   
where $\Phi_{(A)}(\tau, \sigma)$ represents $X^{a\da}_{(A)}$,
$\Psi^{\al\da}_{(A)}$ and $\bar{\Psi}^{\dal \da}_{(A)}$.
We should take the projection onto the $S_{M}$-invariant
subspace.


Because an arbitrary permutation can be decomposed by cyclic permutations,
it is standard to label each twisted sector by a 
Young tableau $(N_1,\ldots,N_l)$, in which each row corresponds 
to the $\bz_{N_i}$-twisted sector.
In the context of $AdS_3/CFT_2$ correspondence, (see, for example,
\cite{MS}) each $\bz_{N_i}$-twisted sector describes 
a single-particle state and $(N_1,\ldots,N_l)$-twisted sector 
describes the Hilbert space of $l$-particle states.
In order to compare with the physical spectrum 
of first quantized superstring theory, it is enough to focus 
on the single-particle Hilbert space.
Hence, we shall concentrate on the $\bz_{N}$-twisted sector 
$(N\leq M)$ from now on.

We label the objects in this sector by the index 
${\cal A}\equiv  \left[(A_0),\ldots ,(A_{N-1}) \right]$, where
$A_i=0,\ldots,N-1$ such that $A_i\neq A_j$ for  ${}^\forall i\neq j$.
The string coordinates of this sector are defined on the world-sheet
$0\leq \sigma \leq 2\pi $, which is rescaled from the
$N$-times one $0\leq \sigma \leq 2\pi N $.
These coordinates are given by
\begin{eqnarray}
&&\Phi_{{\cal A}}(\tau, \sigma) = \Phi_{(A_r)}(\tau, N\sigma - 2\pi r) ~,\nn
&&\frac{2\pi r}{N} \leq \sigma \leq \frac{2\pi (r+1)}{N} ~,~
~r=0,1,\ldots, N-1\,  ~,
\label{long string}
\end{eqnarray} 
where $\Phi$ represents the fields $X^{a\da}$, $\Psi^{\al\da}$ and 
$\bar{\Psi}^{\dal\da}$.
These variables 
$X^{a\da}_{({\cal A})}$, $\Psi^{\al\da}_{({\cal A})}$ and 
$\bar{\Psi}^{\dal\da}_{({\cal A})}$ can be used to construct
the $\cN =4$ superconformal currents 
$\{L_{(\cA),n},G_{(\cA), r}^{\al a}, K_{(\cA), n}^{\al\beta}\}$ 
with  the central charge $c=6$ in the manner similar to  \eqn{SCA}.
However, we have to impose the $\bz_N$-invariance condition 
on the physical Hilbert space as 
\begin{equation}
L_{(\cA),\, 0}- \bar{L}_{(\cA),\, 0} \in N\bz~.
\label{ZN}
\end{equation}
The superconformal currents compatible with this condition \eqn{ZN},
which can act on the physical Hilbert space {\em independently 
of the right movers\/},
consists only of the modes of $n \in N\bz$\footnote
     {The fractional modes, e.g.,
$\hat{L}_{\frac{n}{N}}\equiv
\frac{1}{N}L_{n}$  (where $n\not\in N\bz$), 
can also act on the physical Hilbert space. 
However, in that case, the left and right movers are not independent 
and restricted by the condition \eqn{ZN}.}. 
More precisely, 
the superconformal currents 
describing properly the $\bz_N$-twisted sector 
$\{\hat{L}_n,\hat{G}_r^{\al a}, \hat{K}^{\al\beta}_n\}$ 
should be defined as follows \cite{FKN,BHS} 
(from now on, we shall omit the label $\cA$ for simplicity and 
only present the NS sector)
\begin{eqnarray}
&& \hat{L}_n = \frac{1}{N} L_{nN} + \frac{N^2-1}{4N}\delta_{n0} ~,   \nn
&& \hat{G}_r^{\al a} =\left\{\begin{array}{ll}
               \frac{1}{\sqrt{N}} G_{Nr}^{\al a \,(NS)}~, &  (N=2q+1) ~, \\
	       \frac{1}{\sqrt{N}} G_{Nr}^{\al a \,(R)}~, &  (N=2q) ~,
  	\end{array} \right.  \nn
&& \hat{K}^{\al\beta}_n = K^{\al\beta}_{nN} ~.
\label{hat}
\end{eqnarray}
One can check directly that 
these operators  generate  the $\cN=4$ superconformal algebra with $c=6N$.
The anomaly term in the expression of $\hat{L}_n$ 
corresponds essentially to the Schwarzian derivative of the conformal 
mapping $z~\longmapsto ~ z^N$. 
We should note that the modes of hatted currents are counted by $\hat{L}_0$.

{}From these definitions \eqn{hat} we can find that 
the vacuum $\ket{0;N}$ of $\bz_N$-twisted sector  
possesses the following properties:
\begin{itemize}
 \item $N=2q+1$
\begin{eqnarray}
&& \hat{L}_0\ket{0;N}= \frac{N^2-1}{4N} \ket{0;N} ~,  \nn
&& \hat{K}^3_0\ket{0;N} = 0 ~.
\end{eqnarray}
 \item $N=2q$
\begin{eqnarray}
&& \hat{L}_0\ket{0;N}= \left(\frac{N^2-1}{4N}
+\frac{1}{4N} \right)\ket{0;N} 
\equiv \frac{N}{4}\ket{0;N} ~,  \nn
&& \hat{K}^3_0\ket{0;N}= -\frac{1}{2} \ket{0;N} ~.
\end{eqnarray}
\end{itemize}  
When $N$ is even, the supercurrent 
$\hat{G}_r^{\al a}$ in the  NS sector is made of the one in the R sector
before imposing the $\bz_N$-invariance. 
The extra vacuum energy ``$ \frac{1}{4N}$''
and the extra R-charge  ``$-\frac{1}{2}$'' 
in the case of $N=2q$ originate from this fact.

Now, we focus on the BPS states (chiral primary states)
in this $\bz_N$-twisted sector, which are defined by the next conditions 
\begin{eqnarray}
&& \hat{L}_n \ket{\al}=\bar{\hat{L}}_n \ket{\al}=0 ~,~~
({}^\forall n \geq 1) ~, \nn
&&  \hat{G}^{+ a}_r \ket{\al} = \bar{\hat{G}}^{+ a}_r \ket{\al}=0 ~,~~ 
 ({}^\forall r \geq -\frac{1}{2}) ~, \nn
&&  \hat{G}^{- a}_r \ket{\al} = \bar{\hat{G}}^{- a}_r \ket{\al}=0 ~,~~ 
({}^\forall r \geq \frac{1}{2}) ~.
\label{hat CP}
\end{eqnarray}
These conditions lead inevitably to
\begin{equation}
 (\hat{L}_0-\hat{K}^3_0)\ket{\al}=0 ~.
\end{equation}

At this point it is not difficult to present the explicit forms 
of all the possible BPS states in the $\bz_N$-twisted sector.
(See, for example, \cite{Sugawara}.)
They are written as
\begin{equation}
\ket{\om^{(q,\bar{q})};N}= \ket{\om^q; N }\otimes
\overline{\ket{\om^{\bar{q}};N}} ~, ~~{}^{\forall}\om^{(q,\bar{q})} \in 
H^{q,\bar{q}} (T^4) ~, ~~(q,\bar{q}=0,1,2) ~,
\end{equation}
where the left(right)-moving parts are defined by
\begin{itemize}
\item $N=2q+1$
\begin{eqnarray}
&&  \ket{\om^0;N}= \prod_{i=0}^{q-1}\,
\Psi^{++}_{-\left(\frac{1}{2N}+\frac{i}{N} \right)}
\Psi^{+-}_{-\left(\frac{1}{2N}+\frac{i}{N} \right)} \ket{0;N} ~, \nn
&&  \ket{\om^{1\da};N}= \Psi^{+\da}_{-\frac{1}{2}}
\ket{\om^0;N} ~, \nn
&&  \ket{\om^{2};N}= \Psi^{++}_{-\frac{1}{2}}
\Psi^{+-}_{-\frac{1}{2}}
\ket{\om^0;N} ~. 
\label{cp 1}
\end{eqnarray}
\item $N=2q$
\begin{eqnarray}
&& \ket{\om^0;N}= \prod_{i=0}^{q-1}\,
\Psi^{++}_{-\frac{i}{N}}
\Psi^{+-}_{-\frac{i}{N}} \ket{0;N} ~, \nn
&&  \ket{\om^{1\da};N}= \Psi^{+\da}_{-\frac{1}{2}}
\ket{\om^0;N} ~, \nn
&&  \ket{\om^{2};N}= \Psi^{++}_{-\frac{1}{2}}
\Psi^{+-}_{-\frac{1}{2}}
\ket{\om^0;N} ~. 
\label{cp 2}
\end{eqnarray}
\end{itemize}
It is easy to check that
\begin{eqnarray}
&& \hat{L}_0 \ket{\om^{(q,\bar{q})};N} = \hat{K}^3_0 
\ket{\om^{(q,\bar{q})};N} = 
\frac{q+N-1}{2} \ket{\om^{(q,\bar{q})};N} ~, \nn
&&  \bar{\hat{L}}_0 \ket{\om^{(q,\bar{q})};N} = \bar{\hat{K}}^3_0 
\ket{\om^{(q,\bar{q})};N} =
 \frac{\bar{q}+N-1}{2} \ket{\om^{(q,\bar{q})};N}  ~.
\label{cp spectrum}
\end{eqnarray}


\subsection{``PP Wave Limit'' of $Sym^{M}(T^4)$ $\sigma$-Model and
Comparison with the Superstring Spectrum}
\indent

Now, we focus on the (almost) BPS states with large R-charges. 
As we observed in \eqn{cp spectrum}, 
$\hat{L}_0+\hat{K}^3_0$ takes the value of order $\cO(N)$
for the BPS states in the $\bz_N$-twisted sector. 
In the context of $AdS_3/CFT_2$ correspondence, $\hat{L}_0$ and 
$\hat{K}^3_0$ are identified with $-J^3_0$  and  $K^3_0$, respectively,
which are the zero-modes of the total currents in $SL(2;\br)$ and 
$SU(2)$ super WZW models. Recalling the relations \eqn{contraction},
we must take the identification
\begin{equation}
 \cJ ~\longleftrightarrow~ \hat{K}^3_0-\hat{L}_0~,~~~
 \cF ~\longleftrightarrow~ \frac{1}{k} (\hat{K}^3_0+\hat{L}_0)~.
\end{equation}
Therefore we should consider the case in which $N\sim k \gg 1$. 
We should also assume that $|\hat{K}^3_0-\hat{L}_0| \ll k$. 
In this situation, it is very useful to introduce the next ``pp-wave
limit''.  We redefine 
\begin{eqnarray}
&& \cB^{\pm\da \,'}_n = \Psi^{\pm\da}_{\mp\frac{1}{2}+\frac{nk}{N}}~,~~~
\cA^{a\da\, '}_n = \frac{1}{\sqrt{N}}\,i\partial X^{a\da}_{\frac{nk}{N}}~,\nn
&& \cP^{(l)\,'}_{1,\,n} = -\frac{1}{\sqrt{N}}
\left\{\frac{N}{nk+l}\hat{L}_{\frac{nk+l}{N}}
-\left(\frac{N}{nk+l}-1\right)\hat{K}^3_{\frac{nk+l}{N}}\right\}~,\nn
&& \cP^{*(l)\,'}_{1,\,n} = \frac{1}{\sqrt{N}}
\left\{\frac{N}{nk-l}\hat{L}_{\frac{nk-l}{N}}
-\left(\frac{N}{nk-l}+1\right)\hat{K}^3_{\frac{nk-l}{N}}\right\}~,\nn
&& \cP^{(l)\,'}_{2,\,n} = \frac{1}{\sqrt{N}} \hat{K}^+_{-1+\frac{nk+l}{N}}~,
~~~
\cP^{*(l)\,'}_{2,\,n} = \frac{1}{\sqrt{N}} \hat{K}^-_{1+\frac{nk-l}{N}}~, 
\nn
&& \cQ^{++\pm(l)\,'}_n =\pm \frac{\sqrt{N}}{nk+l}
\hat{G}^{+\pm}_{-\frac{1}{2}+\frac{nk+l}{N}}~, ~~~
\cQ^{--\pm(l)\,'}_n =\mp \frac{\sqrt{N}}{nk-l}
\hat{G}^{-\pm}_{\frac{1}{2}+\frac{nk-l}{N}}~, \nn
&& \cQ^{+- \pm \,'} = \pm \hat{G}^{- \pm}_{1/2}~, ~~~ 
\cQ^{-+ \pm \,'} = \pm \hat{G}^{+ \pm}_{-1/2}~, 
\label{DDF symmetric}
\end{eqnarray}
where $l$ runs over the range $l=1,2,\ldots, k-1$, and take the large
$k$ limit with keeping $\cF$ finite.
Remarkably, these operators satisfy the same (anti-)commutation
relations as \eqn{DDF comm 1} and \eqn{DDF comm 2} with the
identification $N=pk+l$ and $\eta = l/k$ under the large $k$ limit.
Strictly speaking, the mode expansions of twisted $\sigma$-model coordinates
$i\partial X^{a\da}$ and  $\Psi^{\al\da}$ depend on whether $N$ is even or
odd. (Recall the discussions in the previous subsection.)
However, we can neglect safely the difference $1/2N$ under the
assumption of large $N$. 
It is quite important to note that there is the equal number of degrees 
of freedom after taking such large $N$ limit.  
The Hilbert space of $\bz_N$-twisted sector is spanned by 
the free oscillators $\Psi^{\pm\da}_{\mp\frac{1}{2}+\frac{n}{N}}$ and
$i\partial X^{a\da}_{\frac{n}{N}}$.
For each energy level, there are the equal number of 
bosonic and fermionic oscillators defined in \eqn{DDF symmetric}.

At this stage, we can compare the spectra of BPS and almost BPS states 
with the superstring spectrum. As already mentioned, the spectrum
should be compared with the short string sectors with positive $p\geq 1$.
We again forget the sectors with momenta along $T^4$ sector for the time
being.  

First, we consider the BPS states. 
We start with the simple consideration of degrees of freedom. 
At least with respect to the BPS states, 
we can expect that the equal number of physical states exist  
in both of the superstring theory on $AdS_3 \times S^3 \times T^4$ 
and the pp-wave background. This statement is valid  
as long as the pp-wave string theory is defined by the contraction
(\ref{contraction}). 
Fix the spectral flow number $p(\geq 1)$.  
It is known \cite{HHS,AGS} that, roughly speaking, there are about
$k\times \dim H^*(T^4)$ BPS states with the R-charges
$\frac{1}{2}kp \lsim Q  \lsim \frac{1}{2}k(p+1)$ in the 
string theory on the $AdS_3 \times S^3 \times T^4$.
This values of R-charges amount to 
$\cF = p + l/k$ ($0\leq l \leq k$), 
for each of the spectrally flowed sector.    
Therefore, we should assume $\eta = l/k$ $(l=1,2,\ldots,
k-1)$, so that the following physical Hilbert space
\begin{equation}
 \cH_{\msc{pp-wave}}^{(p)} = 
 \bigoplus_l  \cH_{\msc{pp-wave}}(p, \eta=l/k)
\label{hilbert pp}
\end{equation}
includes the equal number of BPS states as those of  $AdS_3\times S^3
\times T^4$ superstring theory\footnote
   {The threshold values  $l=0$ and $k$ cannot be included 
    because there is a restriction $0<\eta<1$ in the type II
    representations.  Moreover, the states with
    $l=k-1$ corresponds to the missing states in the original   
    $AdS_3\times S^3$ string theory for each $p$. 
    (See, for example, \cite{HS2,HHS}.)
    One might worry that $\eta$ should be continuous in principle. 
    These subtleties are, however, harmless for sufficiently large $k$.}.

Turning to the symmetric orbifold, we consider the following direct
sum of the single particle Hilbert spaces of the $\bz_{N(l)}$ 
twisted sectors, (where we set $\dsp N(l)\equiv kp+l$ $(0< l < k)$)
\begin{equation}
 \cH_{\msc{symm}}^{(p)} = 
 \bigoplus_l  \cH_{\msc{symm}}(N(l)= kp+l) ~.
\label{hilbert symm}
\end{equation}
It is obvious that $\cH_{\msc{symm}}^{(p)}$ and 
$\cH_{\msc{pp-wave}}^{(p)}$ have the equivalent 
spectrum of BPS states. 
We also point out that the identification
$\cB^{+\da'}_0 (\equiv \Psi^{+\da}_{-\frac{1}{2}}) = \cB^{+\da}_0$ 
is consistent with the relation between 
\eqn{BPS relation} and \eqn{cp 1}, \eqn{cp 2}.

Next, we consider the almost BPS states.
Once we admit the correspondence of $\eta = l/k$, 
it is not difficult to find the almost BPS states in 
$\cH_{\msc{symm}}^{(p)}$ that correspond to the stringy excitations
of the string theory side. 
First, we rewrite the DDF operators  \eqn{DDF1} and \eqn{DDF2}, which 
act on the Hilbert space $\cH_{\msc{pp-wave}}(p, \eta=l/k)$,
as $\cP^{(l)}_{i,\,n}$, $\cP^{*(l)}_{i,\,n}$ and so on.
We also rewrite the operators \eqn{DDF symmetric} as 
$\cP^{(l,j)'}_{i,\,n}$, $\cP^{*(l,j)'}_{i,\,n}$ and so on,
for each $\bz_{N(j)}$-twisted sector $\cH_{\msc{symm}}(N(j))$
(where $1\leq l \leq k-1$ and $1\leq j \leq k-1$).
The ``diagonal terms'' (e.g., $\cP^{(l,l)'}_{i,\,n}$) generate 
the superalgebra equivalent to that of DDF operators 
$\cP^{(l)}_{i,\,n}$ , $\cP^{*(l)}_{i,\,n}$ and so on, as we already
mentioned. Now the correspondence is obvious.
The non-trivial consistency check is only the level matching condition. 
In the symmetric orbifold theory side, the level matching condition 
is given by
\begin{equation}
\hat{L}_0-\bar{\hat{L}}_0 \in \bz~.
\end{equation} 
This condition is consistent with 
the result of string theory side \eqn{lm string} under the above 
correspondence of DDF operators.

However, we should note here that the string Hilbert space 
$\cH_{\msc{pp-wave}}^{(p)}$ is strictly smaller than that of 
symmetric orbifold; that is to say
\begin{equation}
\cH_{\msc{pp-wave}}^{(p)} \stackrel{\subset}{\neq}
\cH_{\msc{symm}}^{(p)}~. 
\end{equation} 
In fact, the ``non-diagonal terms''(e.g., $\cP^{(l,j)'}_{i,\,n}$
with $l\neq j$) have no counterparts in the string side. 
Therefore, we cannot define the corresponding DDF operators as local
operators on the string Hilbert space.
These missing states in the string spectrum may be compensated by 
non-perturbative excitations. 
Because we are now assuming the small string coupling, 
the non-perturbative excitations usually become very massive.
Under the assumption of large $k$,
the space of almost BPS states can include such very massive 
excitations in principle. However, our world-sheet analysis 
as a perturbative string theory cannot include such excitations. 
This is the reason why there are many missing states 
in the string side.
The non-perturbative analysis will be an important task for future study.

Finally we consider the sectors with non-vanishing momenta.
As in the string spectrum, there are no BPS states in these sectors,
however there are many almost BPS states. 
Recall that $\cJ= \hat{K}^3_0 -\hat{L}_0$ and
$\hat{L}_0 = \frac{1}{N} L_{0} + \frac{N^2-1}{4N}$.
The operator $L_0$ includes the contribution of momenta with the standard 
normalization. Hence, we find that the contributions of momenta
to $\cJ$ and $\bar{\cJ}$ are given as follows (for the sector 
$\cH_{\msc{symm}}(N(l))$ with $N(l)\equiv kp +l$)
\begin{eqnarray}
&& \Delta \cJ = -\frac{1}{2N(l)} \sum_a
\left(\frac{n'_a}{R_a}+\frac{w'^aR_a}{2}\right)^2~,~~
\Delta \bar{\cJ} = -\frac{1}{2N(l)} \sum_a
\left(\frac{n'_a}{R_a}-\frac{w'^aR_a}{2}\right)^2~,
\end{eqnarray}
where $n_a'$ and $w'^a$ are KK momenta and winding modes as before.
The level matching condition for the vacuum state can be read as
\begin{equation}
\sum_a n'_aw'^{a} \in N(l)\bz~.
\end{equation}
By comparing the analysis given in the last section (under the  
identification $\eta = l/k$),
we obtain the following correspondence between the spectrum of string theory 
and that of the symmetric orbifold theory as
\begin{equation}
   n'_a= \sqrt{k}n_a~,~~~ w'^{a} = \sqrt{k}w^a~, 
\end{equation}
where $n_a$ and $w^a$ are the zero-momenta in the string theory
side\footnote{
   Although $\sqrt{k}$ is not an integer in general, 
   we can approximate it by $\lb \sqrt{k} \rb$, because we are now 
   assuming  large $k$, such as 
   $|\sqrt{k}-\lb \sqrt{k}\rb|/\sqrt{k} \ll 1$.
    }.

We have observed that there are again many missing states in the string
theory side. 
The essentially same aspect was already pointed out in the context of 
string theory on $AdS_3\times S^3 \times T^4$ \cite{KLL}.
It may be worthwhile to comment on how such discrepancy is removed  
if assuming the fractional string excitations. 
These excitations do not exist in the perturbative string spectrum and
may be at least explained in the S-dual picture. 
The existence of $k\equiv Q_5$ NS5
leads to the fractional string with the tension 
$ \tilde{T} = T / k$, where $T$ represents the tension of 
fundamental string. If we measure the radii of $T^4$ by the unit of
string length $l_s = 1/\sqrt{T}$, namely, $R_a = r_a l_s$,
the momenta \eqn{T4 momenta} becomes
\begin{equation}
p_a = \frac{1}{l_s}\left(\frac{n_a}{r_a}+w^ar_a\right)  ~,~~~
\bar{p}_a =\frac{1}{l_s}\left(\frac{n_a}{r_a}-w^ar_a\right) ~.
\label{T4 momenta 2}
\end{equation} 
For the fractional strings, we should replace the string length $l_s$
with $\tilde{l}_s \equiv \sqrt{k} l_s$. Thus, we obtain
\begin{equation}
p_a = \frac{1}{\sqrt{k}}\times
\frac{1}{l_s}\left(\frac{n_a}{r_a}+w^ar_a\right)  ~,~~~
\bar{p}_a = \frac{1}{\sqrt{k}}\times
\frac{1}{l_s}\left(\frac{n_a}{r_a}-w^ar_a\right) ~,
\label{T4 momenta 3}
\end{equation}
and the extra factor $1/\sqrt{k}$ completely compensates  the spectrum of 
missing states.

~

\section{Extension to the Case of $H_6  \times T^4/\bz_2$}

\indent 

In this section, we extend our previous analysis to the case of 
superstring theory on $H_6  \times T^4/\bz_2$ background and 
the symmetric orbifold $Sym^{M}(T^4/\bz_2)$. 
We again find the good correspondence for the (almost) BPS 
spectrum.

\subsection{Spectrum of Superstring on $H_6  \times T^4/\bz_2$}
\indent

The $\bz_2$-orbifold action acts on the string coordinates as 
\begin{equation}
\la^{a\da} ~\longrightarrow~ -\la^{a\da}~, ~~~
Y^{a\da}~ \longrightarrow~ -Y^{a\da}~.
\label{Z2 orbifold 1}
\end{equation}
Moreover, we assume the action on the free bosons $H^3$ and $H^4$,
which are bosonizations of $\la^{a\da}$, as
\begin{equation}
 H_3 ~\longrightarrow~H_3+\pi~,~~~ H_4~\longrightarrow~H_4-\pi~,
\label{Z2 orbifold 2}
\end{equation}
so that the space-time supersymmetry is preserved. 
In fact, we can directly check that all the generators of
super pp-wave algebra $\{\cJ, \cF, \cP_i, \cP_i^*, \cQ^{\al\beta a}\}$ 
defined in \eqn{PVF} and \eqn{PVS} are invariant 
under \eqn{Z2 orbifold 1} and \eqn{Z2 orbifold 2}.
As for the DDF operators \eqn{DDF1} and \eqn{DDF2},
the orbifold action reads as 
\begin{equation}
\cA^{a\da}_n ~ \longrightarrow~ - \cA^{a\da}_n~, ~~~
\cB^{\al\da}_n ~ \longrightarrow~ - \cB^{\al\da}_n~,
\end{equation}
and the other DDF operators are invariant under this action.

~

Now, we write down the spectrum of (almost) BPS states.

~

\noindent{\bf 1.  Untwisted sector}

The analysis of the untwisted sector is quite easy.
First, we consider the BPS states.
All the task we have to do is 
to leave the $\bz_2$-invariant BPS states in the case of $T^4$. 
Only the NS-NS and R-R BPS states are left and the NS-R and 
R-NS BPS states are projected out.
Thus, we obtain 8 BPS states for each $p$ and $\eta$, 
and they are identified with the even cohomology of $T^4$.
As for the almost BPS states, we first consider the following Fock vacua as
\begin{equation}
\ket{i,\bar{i};j,\bar{j};\eta,p;n_a,w^a; (\pm)} =
\ket{i,\bar{i};j,\bar{j};\eta,p;n_a,w^a} \pm 
\ket{i,\bar{i};j,\bar{j};\eta,p;-n_a,-w^a} ~,
\end{equation}
where $\ket{i,\bar{i};j,\bar{j};\eta,p;n_a,w^a}$ represents the physical 
state with the non-vanishing momenta along $T^4$, which was defined in
\eqn{phys T4 momenta}.   Clearly,  the (NS-NS) BPS states 
are realized as the special cases of such physical states; that is
$\ket{i,\bar{i};0,0;\eta,p;0,0; (+)}$. (R-R BPS states can be obtained 
by multiplying $\cB^{+\da}_0$ and $\bar{\cB}^{+\da}_0$.)
The general physical states 
are constructed  by multiplying the DDF operators and supercharges 
$\cQ^{\pm\mp a}$ as in the case of $T^4$. 
We only need the following additional constraint as
\begin{eqnarray}
&& \sharp \{\cB_{-n}^{\al\da},~ \cA_{-m}^{a\da}\} + 
   \sharp \{\bar{\cB}_{-n}^{\al\da},~ \bar{\cA}_{-m}^{a\da}\} = \mbox{even}~,
   ~\left( \mbox{for the Fock vacua $\ket{i,\bar{i};\cdots ; (+)}$} 
       \right)~, \nn
&& \sharp \{\cB_{-n}^{\al\da},~ \cA_{-m}^{a\da}\} + 
   \sharp \{\bar{\cB}_{-n}^{\al\da},~ \bar{\cA}_{-m}^{a\da}\} = \mbox{odd}~,
   ~\left( \mbox{for the Fock vacua $\ket{i,\bar{i};\cdots ; (-)}$}
         \right) ~.
\end{eqnarray}

~

\noindent{\bf 2.  Twisted sectors}

There are 16 twisted sectors that describe stringy excitations around 
each fixed point of orbifold action. 
For each twisted sector,  
we need to consider the following boundary conditions as
\begin{eqnarray}
&& Y^{a\da}(e^{2\pi i}z ) = - Y^{a\da}(z ) ~,\nn
&& \la^{a\da}(e^{2\pi i}z ) = - \la^{a\da}(z )  ~,~
                                      (\mbox{for  NS sector}) ~,\nn
&& \la^{a\da}(e^{2\pi i}z ) = \la^{a\da}(z )  ~,~
                                      (\mbox{for R sector}) ~, 
\end{eqnarray}
and moreover,
\begin{equation}
H_3(e^{2\pi i}z) = H_3(z) + \pi~, ~~~ H_4(e^{2\pi i}z) = H_4(z) - \pi~.
\label{twisted H34}
\end{equation}

First, we consider the BPS states. In the NS vacua, there are  both of 
the bosonic and fermionic twist fields; $\sigma_{\bsz_2}^{b}$, 
$\sigma_{\bsz_2}^{f}$, whose conformal weights are equal to
\begin{equation}
h(\sigma_{\bsz_2}^{b})= \bar{h}(\sigma_{\bsz_2}^{b})
= 4\times \frac{1}{16}=\frac{1}{4}~,~~~
h(\sigma_{\bsz_2}^{f})= \bar{h}(\sigma_{\bsz_2}^{f})
= 4\times \frac{1}{16}=\frac{1}{4}~.
\end{equation}
Based on this fact we can observe that there are no BPS
states in the NS-NS, NS-R and R-NS sectors. 
On the other hand, the R vacua can include only the bosonic twist field
$\sigma_{\bsz_2}^b$ and only the spin fields along  the $H_6$
direction, which we express as $S^{\ep_0\ep_1\ep_2}$.
This fact leads us to a unique R-R BPS state (per each twisted sector),
which is explicitly written as 
\begin{equation}
S^{-++}_{-\frac{3}{8}+\eta}\bar{S}^{-++}_{-\frac{3}{8}+\eta}
\sigma_{\bsz_2}^b\ket{0,\eta,p}\otimes 
c\bar{c}e^{-\frac{\phi}{2}-\frac{\bar{\phi}}{2}}\ket{0}_{\msc{gh}}~. 
\label{twisted BPS}
\end{equation}
In this way, we have found the 16 R-R BPS states in the twisted sectors
for each $p$ and $\eta$.
They correspond to the blow-up modes of $T^4/\bz_2$ orbifold
and reproduce the cohomology ring of $K3$ together with the
contributions from the untwisted sector.

The other physical states are also straightforwardly constructed. 
Contrary to the untwisted sector, the states with non-trivial
momenta along $T^4$ are not allowed. Thus, we only have to
consider the states created by the actions of DDF operators over the 
BPS states \eqn{twisted BPS}. The only non-trivial point is that 
the modes of DDF operators $\cB^{\al\da}_r$ should be half integers
$ r\in \frac{1}{2}+\bz$ in this case. This fact originates from 
the boundary condition \eqn{twisted H34}. We again need the constraint
\begin{equation}
\sharp \{\cB_{-r}^{\al\da},~ \cA_{-m}^{a\da}\} + 
   \sharp \{\bar{\cB}_{-r}^{\al\da},~ \bar{\cA}_{-m}^{a\da}\} = \mbox{even}~,
\end{equation}
to preserve the $\bz_2$-invariance.


\subsection{Comparison with $Sym^M(T^4/\bz_2)$ $\sigma$-Model}
\indent

The $\bz_2$-orbifoldization of $Sym^M(T^4)$ $\sigma$-model is 
defined by the action
\begin{equation}
 X_{(A)}^{a\da} ~ \longrightarrow ~ - X_{(A)}^{a\da}~, ~~~
 \Psi^{\al\da}_{(A)} ~ \longrightarrow ~ - \Psi^{\al\da}_{(A)}~.
\end{equation} 
This action preserves the $\cN=(4,4)$ superconformal symmetry.
We again study the single particle Hilbert space of $\bz_N$-twisted
sector with $N =pk+l$ ($0<l<k$), as in the previous section.

We first consider the spectrum of BPS states.
In the untwisted sector of $\bz_2$ orbifoldization, 
only the $\bz_2$-invariant BPS states survive. 
They correspond to the even cohomology of $T^4$. 

The analysis of the twisted sectors is more complicated. 
Focusing on one of the twisted sectors corresponding to the 16 fixed
points, we can observe the following aspects:
\begin{itemize}
 \item $N=2q+1$

We have the mode expansions 
$i\partial X^{a\da}_{\frac{n}{N}+\frac{1}{2N}}$ and
$\Psi^{\al\da}_{\frac{n}{N}}$, ($n\in \bz$) and we obtain
for the NS vacuum $\ket{0;N}^{\msc{(t)}}$
\begin{eqnarray}
&& \hat{L}_0\ket{0;N}^{\msc{(t)}} 
= \left(\frac{N^2-1}{4N}+\frac{1}{2N}\right)\ket{0;N}^{\msc{(t)}}
\equiv \frac{N^2+1}{4N}\ket{0;N}^{\msc{(t)}} ~, \nn
&& \hat{K}^3_0\ket{0;N}^{\msc{(t)}}
 = -\frac{1}{2}\ket{0;N}^{\msc{(t)}}~.
\end{eqnarray}
 \item $N=2q$

We have the mode expansions 
$i\partial X^{a\da}_{\frac{n}{N}+\frac{1}{2N}}$ and
$\Psi^{\al\da}_{\frac{n}{N}+\frac{1}{2N}}$, ($n\in \bz$) and we obtain
for the NS vacuum $\ket{0;N}^{\msc{(t)}}$
\begin{eqnarray}
&& \hat{L}_0\ket{0;N}^{\msc{(t)}}
 = \left(\frac{N^2-1}{4N}+\frac{1}{4N}\right)\ket{0;N}^{\msc{(t)}}
\equiv \frac{N}{4}\ket{0;N}^{\msc{(t)}} ~, \nn
&& \hat{K}^3_0\ket{0;N}^{\msc{(t)}} = 0~.
\end{eqnarray}
\end{itemize}
In these expressions the extra zero-point energies and R-charges
assigned to the NS vacua are due to these twisted mode expansions.
Based on these aspects, we can find out the following BPS state that is 
unique for each of the twisted sectors as
\begin{eqnarray}
&&\ket{\om^{(1,1)};N}^{\msc{(t)}}=
 \ket{\om^{1};N}^{\msc{(t)}} 
\otimes \overline{\ket{\om^{1};N}}^{\msc{(t)}} ~,\\
&& \ket{\om^1;N}^{\msc{(t)}}  
= \prod_{i=0}^{q}\,\Psi^{++}_{-\frac{i}{N}}\Psi^{+-}_{-\frac{i}{N}}
  \ket{0;N}^{\msc{(t)}}~, ~~~(N=2q+1) ~, \nn
&& \ket{\om^1;N}^{\msc{(t)}}  
= \prod_{i=0}^{q-1}\,\Psi^{++}_{-\frac{i}{N}-\frac{1}{2N}}
   \Psi^{+-}_{-\frac{i}{N}-\frac{1}{2N}}
  \ket{0;N}^{\msc{(t)}}~, ~~~(N=2q) ~,
\end{eqnarray}
and we obtain 
\begin{equation}
\hat{L}_0\ket{\om^1;N}^{\msc{(t)}}=
\hat{K}^3_0\ket{\om^1;N}^{\msc{(t)}}
= \frac{N}{2}\ket{\om^1;N}^{\msc{(t)}}~.
\end{equation}

In summary, we have obtained $8+16=24$ BPS states, which precisely
correspond to the cohomology of $K3$ for each of the $\bz_N$-twisted
sectors.  They have (approximately) degenerate charges
$  \cF (\equiv (\hat{K}^3_0+\hat{L}_0) /k ) =
p+l/k$, 
and $\cJ (\equiv \hat{K}^3_0-\hat{L}_0)=0$.
As in the case of $T^4$,  
we have a good correspondence between the string theory and symmetric
orbifold theory under the identifications $N=kp+l$ and $\eta=l/k$.

With respect to the almost BPS states, the discussion is almost parallel 
to the case of $T^4$.  
We can explicitly write down the complete list of almost BPS states in
the symmetric orbifold theory and compare it with the string theory result 
by using the identification of DDF operators, as before.
However, in this case, we must identify the DDF operators 
$\cB^{\al\da}_{n+\frac{1}{2}}$  (where $n \in \bz$) in the twisted sectors
with $\Psi^{\pm\da}_{\mp\frac{1}{2}+\frac{nk}{N}+\frac{k}{2N}}$
rather than $\Psi^{\pm\da}_{\mp\frac{1}{2}+\frac{nk}{N}}$.  
(We again neglect the small difference of mode $1/2N$.)
The short string spectrum is again completely 
embedded in the Hilbert space of symmetric orbifold theory and there are 
many missing states.

~

\section{Conclusion}
\indent

In this paper we have studied the correspondence between string theory on 
pp-wave background with NSNS-flux and  superconformal theory on
symmetric orbifold $Sym^{Q_1Q_5}(M^4)$,
where $M^4 = T^4$ or $T^4/\bz_2 (\cong K3)$.
Superstring theory on the pp-wave background is obtained as
the Penrose limit of superstring on the $AdS_3 \times S^3 \times M^4$
with NSNS-flux \cite{Sfetsos}. 
This theory is described by a noncompact WZW model 
with the target manifold of 6-dimensional Heisenberg group ($H_6$).
We employed current algebra approach according to \cite{KK}
and quantize the system in the covariant gauge.
By making use of the free field representation, 
we have explicitly constructed the physical vertices that correspond to the
generators of global supersymmetries and the complete set of DDF operators.
The spectrum of physical states is classified by the ``short string
sectors'' and ``long string sectors'', as in the $AdS_3$ string 
theory \cite{MO}.  
The latter has a continuous spectrum of light-cone energy
and the strings freely propagate 
along the transverse plane, while the strings of the former 
cannot propagate along the transverse plane.

We have compared the general short string excitations 
with the single particle Hilbert spaces included  
in the symmetric orbifold theory.  We have shown that all the 
the short string states are successfully 
embedded into the Hilbert space of symmetric orbifold theory.
At this point, our analysis of DDF operators played an essential role. 
We have also found the existence of many missing states in the
string Hilbert space, which should be understood as non-perturbative 
excitations.

It has just begun to study the duality between string theory on pp-wave
backgrounds and boundary conformal field theory.
Among other things, it is a very interesting and challenging problem how we 
should explain the origin of missing states mentioned above. 
For this purpose it may be a suggestive thing 
that our $H_6$ string theory can be described by the conformal theory
reminiscent of $\bc^2/\bz_k$ model under our identification  $\eta = l/k$. 
Roughly speaking, this seems to originate 
from the well-known fact that the $k$ NS5 system 
should be the T-dual of $\bc^2/\bz_k \sim ALE(A_{k-1})$,
although the original $AdS_3 \times S^3$ background is known 
to be located at the different point in the moduli space\footnote
     {The $\bc^2/\bz_k$ model corresponds to the point with
      non-vanishing world-sheet theta angle $\theta = 1/k$, 
      but the $AdS_3 \times S^3$ model 
      should correspond to the point of $\theta=0$. 
      However, the difference become negligible 
      in the large $k$ limit.}.
In any case, the similarity with the $\bc^2/\bz_k$ model seems to imply 
the existence of fractional string excitations, if we take
a non-perturbative approach, say, the matrix string theory \cite{Matrix}.
As we discussed in section 4, such excitations may 
give a good explanation of the missing states.

Another important problem is to determine the interactions  
from the world-sheet techniques by calculating the correlation functions. 
Recently, the interactions in string theory on pp-wave backgrounds
with RR-flux have been investigated \cite{interact}.
It may be interesting to analyse the string interaction in our
NSNS-model and compare it with the results of the RR-background.



\section*{Acknowledgement}
\indent

We  would like to thank T. Takayanagi for useful discussions.
The work of Y. S. is supported in part by a
Grant-in-Aid for the Encouragement of Young Scientists 
($\sharp 13740144$) from the Japanese Ministry of Education, 
Culture, Sports, Science and Technology.

~


\section*{Appendix A ~ Gamma Matrices}
\setcounter{equation}{0}
\def\theequation{A.\arabic{equation}}
\indent

The cocycle factors of spin fields are defined by means of 
the gamma matrices. We here summarize the convention in this paper;
\begin{eqnarray}
 \Gamma_{\pm 0} &=& \sigma_{\pm} \otimes \bone 
      \otimes \bone \otimes \bone \otimes \bone ~,\nn
 \Gamma_{\pm 1} &=& \sigma_3 \otimes \sigma_{\pm} 
      \otimes \bone \otimes \bone \otimes \bone ~,\nn
 \Gamma_{\pm 2} &=& \sigma_3 \otimes \sigma_3 
      \otimes \sigma_{\pm} \otimes \bone \otimes \bone ~,\nn
 \Gamma_{\pm 3} &=& \sigma_3 \otimes \sigma_3
     \otimes \sigma_3 \otimes \sigma_{\pm} \otimes \bone ~,\nn
 \Gamma_{\pm 4} &=& \sigma_3 \otimes \sigma_3 
     \otimes \sigma_3 \otimes \sigma_3 \otimes \sigma_{\pm} ~,
\end{eqnarray}
where we use the usual Pauli matrices as
\begin{equation}
 \sigma_1 = \left(
\begin{array}{rcl}
 0 & 1 \\
 1 & 0 
\end{array}
\right)  ~,~
 \sigma_2 = \left(
\begin{array}{rcl}
 0 & -i \\
 i & 0 
\end{array}
\right) ~,~
\sigma_3 = \left(
\begin{array}{rcl}
 1 & 0 \\
 0 & -1 
\end{array}
\right) ~,
\end{equation} 
and $\sigma_{\pm} = \frac{1}{2}(\sigma_1 \pm i \sigma_2) $.
We use the charge conjugation matrix as
\begin{equation}
 C =  \epsilon \otimes \sigma_1 
     \otimes \epsilon \otimes \sigma_1 \otimes \epsilon ~,~~
 \epsilon = i \sigma_2 ~,
\end{equation}
which has the property as 
\begin{equation}
 C \Gamma_{\mu} C^{-1} = - (\Gamma_{\mu})^{T} ~,~ C^{\dagger} = C^{-1} ~.
\end{equation}
Then the OPEs including spin fields are given as follows
\begin{eqnarray}
 \psi^{\mu} (z) S^A (w) &\sim& \frac{1}{(z-w)^{\frac{1}{2}}}  
   (\Gamma^{\mu})^A_{~B} S^B (w) ~, \nn 
 \psi^{\mu} \psi^{\nu} (z) S^A (w) &\sim& 
 - \frac{1}{z-w} 
   (\Gamma^{\mu \nu})^A_{~B} S^B (w) ~, \nn 
 S^A (z) S^B (w) &\sim& \frac{1}{(z-w)^{\frac{3}{4}}} 
  (\Gamma_{\mu} C)^{AB} \psi^{\mu} (w)~.
\end{eqnarray}

~


\section*{Appendix B ~ Supersymmetry of DDF operators}
\setcounter{equation}{0}
\def\theequation{B.\arabic{equation}}
\indent

Supertransformations of DDF operators under the action of $\cQ^{\pm \mp a}$ 
are given in (\ref{CRPQ}), (\ref{CRP*Q}) and (\ref{CRAQ}).
In this appendix, we write them in explicit forms. 
For DDF operators corresponding to ``affine'' $H_6$ extension, 
they are given by
\begin{eqnarray}
&& {[} \cQ^{-+a} , \cP_{1,n} {]} = 
  \frac{n + \eta}{p + \eta}  
   \cQ^{++a}_n ~,~~
 {[} \cQ^{+-a} , \cP^{*}_{1,n} {]} = 
  \frac{n - \eta}{p + \eta}  
   \cQ^{--a}_n ~, \nn
&& {[} \cQ^{+-a} , \cP_{2,n} {]} = 
  \frac{n + \eta}{p + \eta}  
   \cQ^{++a}_n ~,~~
 {[} \cQ^{-+a} , \cP^{*}_{2,n} {]} = - 
  \frac{n - \eta}{p + \eta}  
   \cQ^{--a}_n ~, \nn
&& \{  \cQ^{+-a} , \cQ^{++b}_n \} = 
 - \epsilon^{ab} \cP_{1,n} ~,~~
 \{  \cQ^{-+a} , \cQ^{--b}_n \} =
 \epsilon^{ab} \cP^{*}_{1,n} ~, \nn
&& \{  \cQ^{-+a} , \cQ^{++b}_n \} = 
 \epsilon^{ab} \cP_{2,n} ~,~~
 \{  \cQ^{+-a} , \cQ^{--b}_n \} = 
 \epsilon^{ab} \cP^{*}_{2,n} ~.
\end{eqnarray}
For DDF operators of $T^4$ sectors, they are given by
\begin{eqnarray}
&& {[} \cQ^{+-a} , \cA^{b \dot{b}}_n {]} = 
  \frac{n}{p + \eta} \epsilon^{ab} 
   \cB^{- \dot{b}}_n ~,~~
 {[} \cQ^{-+a} , \cA^{b \dot{b}}_n {]} = 
  - \frac{n}{p + \eta} \epsilon^{ab}  
   \cB^{+ \dot{b}}_n ~,\nn
&& \{  \cQ^{-+a} , \cB^{- \dot{b}}_n \} = 
 - \cA^{a \dot{b}}_n ~,~~
 \{  \cQ^{+-a} , \cB^{+ \dot{b}}_n \} = 
 - \cA^{a \dot{b}}_n ~.
\end{eqnarray}


\end{document}